\documentclass[twocolumn]{aastex631}

\usepackage{graphicx}
\usepackage{amsmath}
\usepackage{multirow}

\newcommand\teff{$T_\mathrm{eff}$} 
\newcommand\logg{log~\textit{g}} 
\newcommand\kep{\textit{Kepler}} 
\newcommand\cn{[C/N]} 
\newcommand\feh{[Fe/H]}

\newcommand\nfe{[N/Fe]}
\newcommand\mgh{[Mg/H]}

\submitjournal{ApJ}

\begin{document}

\title{[C/N] Ages for Red Giants and their Implications for Galactic Archaeology}

\author[0000-0002-2854-5796]{John D. Roberts}
\affiliation{Department of Astronomy, The Ohio State University \\
140 W 18th Ave \\
Columbus, OH 43210, USA}
\affiliation{Center for Cosmology and Astroparticle Physics, The Ohio State University\\
191 W Woodruff Ave \\
Columbus, OH, 43210, USA}

\author[0000-0002-7549-7766]{Marc H. Pinsonneault}
\affiliation{Department of Astronomy, The Ohio State University \\
140 W 18th Ave \\
Columbus, OH 43210, USA}
\affiliation{Center for Cosmology and Astroparticle Physics, The Ohio State University\\
191 W Woodruff Ave \\
Columbus, OH, 43210, USA}

\author[0000-0001-7258-1834]{Jennifer A. Johnson}
\affiliation{Department of Astronomy, The Ohio State University \\
140 W 18th Ave \\
Columbus, OH 43210, USA}
\affiliation{Center for Cosmology and Astroparticle Physics, The Ohio State University\\
191 W Woodruff Ave \\
Columbus, OH, 43210, USA}

\author[0000-0003-3781-0747]{Liam O. Dubay}
\affiliation{Department of Astronomy, The Ohio State University \\
140 W 18th Ave \\
Columbus, OH 43210, USA}
\affiliation{Center for Cosmology and Astroparticle Physics, The Ohio State University\\
191 W Woodruff Ave \\
Columbus, OH, 43210, USA}

\author[0000-0002-6534-8783]{James W. Johnson}
\affiliation{Carnegie Science Observatories \\
813 Santa Barbara St.\\
Pasadena, CA 91101, USA}

\begin{abstract}
Red giants undergo the first dredge-up, a mixing event that creates a connection between their surface [C/N] and their mass and age. We derive a [C/N]-Age relationship for red giants calibrated on APOGEE DR17 abundances and APOKASC-3 asteroseismic ages. We find that we can use \cn\ to reliably recover asteroseismic ages between 1 and 10 Gyr with average uncertainties of 1.64 Gyr. We find that \cn\ yields concordant ages, with modest offsets, for stars in different evolutionary states. We also find that the \cn-birth mass relationship is robust for luminous giants, and argue that this is an advantage over direct asteroseismology for these stars. We use our ages to infer Galactic birth abundance trends in \feh\ and \mgh\ as a function of position in the Galactic disk. We filter out stars with kinematic or chemical properties consistent with migrators and found the number of migrators to be much lower than expected by standard radial migration prescriptions. The remaining population shows weak chemical evolution trends, on the order of 0.01 dex/Gyr, over the last 10 Gyr across a wide range of radii. 
\end{abstract}

\keywords{Galactic archaeology (2178), Red giant stars (1372), Stellar ages (1581)}

\section{Introduction} \label{sec:intro}

Red giants are instrumental to our understanding of our Galaxy. The red giant phase is relatively long-lived, so they are very numerous. Red giants are also luminous and present across a wide range of ages and birth abundances. Their spectra are information-rich, providing detailed data about fundamental stellar properties and the mixture of heavy elements. Their cool surface temperatures also make them well-suited for infrared spectroscopy. These features make them excellent targets for large-scale galactic surveys such as APOGEE \citep{Majewski2017} and Milky Way Mapper (MWM) \citep{SDSSV}, and \added{among} the best subjects we have for learning about the Milky Way on the galactic scale.  

Age is a crucial stellar property that can be remarkably difficult to infer for stars. Red giants with a wide range of ages and abundances are concentrated in a narrow range of temperatures. Unlike the turnoff, luminosity is not a useful age diagnostic \citep{Soderblom_2010}. Since these stars have completed their main-sequence lifetime, their mass is a useful age diagnostic. Mass can be determined from a surface gravity and a radius. Unfortunately, the tight distribution of the RGB on the HR-diagram means that even small uncertainties in measured parameters can result in large uncertainties for masses and ages \citep{Salaris_2002, Feuillet_2016, Morales_2025}. 

For red giant studies, clusters have long played an important role because the ages and masses of cluster members can be accurately and precisely measured from global studies of the cluster. The Gaia-ESO survey \citep{GaiaESO_1, GaiaESO_2} and the APOGEE cluster surveys \citep{OCCAM} have been fruitful efforts. However, only a small fraction of stars are in clusters, and the well-studied star clusters are not a representative sample of the underlying Galactic populations. To truly trace the formation history of the Milky Way galaxy, a much larger sample of stars is needed. 

Red giants exhibit seismic oscillation patterns related to their radius and density \citep{Tassoul_1980, Ulrich_1986, KjeldsenBedding_1995, Belkacem_2011,apokasc3}. Through observing these oscillations, their mass can be determined. These masses can be translated into ages given composition measurements and appropriate stellar models. These masses, and hence ages, are quite precise, but unfortunately rely on short cadence observations that are costly to apply to large samples.

Fortunately, the surface abundances of RGB stars are altered by mixing, and the mixing signal is tied to stellar mass and composition \added{\citep{Iben_1965}}. Stars are, in general, born with more carbon than nitrogen, so unprocessed regions have a high carbon-to-nitrogen ratio \added{\citep{SolarMixture}}. However, carbon is burned much more rapidly in the CN cycle than nitrogen is, so the interiors of stars hot enough to operate the CN cycle have a low carbon-to-nitrogen ratio. This change is hidden on the main sequence, but when stars evolve to the giant branch, they develop deep surface convection zones. During this first dredge-up (FDU), processed material from the deep interior is mixed into the outer convective regions. Higher mass stars have hotter cores, thus a larger processed region, and develop deeper mixing zones. As a consequence, the degree of C to N processing is mass-dependent \added{\citep{Salaris_2015}}. Metallicity also matters, largely because the convection zone depth depends on abundance. 

Turning \cn\ into an age diagnostic can be done by calibrating the \cn\ to mass relationship stellar models or with samples of stars with already known ages and surface abundances \added{\citep{Masseron_2015, Salaris_2015, Martig_2016}}. However, not all stars have the same \cn\ at birth, and differences in composition, such as varying \feh, can produce different mixing strengths, resulting in variations in post-FDU \cn\ for stars of the same mass \citep{Roberts_2024}. \feh\ can be used to track much of the variation in the birth abundance. For more evolved red giants, such as ones near the upper end of the red giant branch or red clump stars, there are additional complications. Metal-poor red giants undergo extra mixing processes that can contaminate the relationship that is set during the FDU \citep[e.g.][]{Gratton_2000, Shetrone_2019}. Additionally, if ages are derived from the mass, such as asteroseismic ages, the mass loss that occurs at the tip of the RGB can also complicate results. \added{Asteroseismology probes the current mass of a star, and turning this into an age requires a model-dependent treatment of mass loss to reconstruct the main-sequence mass of the star}. However, the \cn-age relationship is set at the base of the branch, prior to this mass loss, allowing it to ignore this issue. However, care must still be taken to ensure this process does not confuse results \added{through altering training datasets.}

The use of \cn\ as an age diagnostic has been explored by many papers in recent years. \citet{Salaris_2015} and \citet{Lagarde_2017} have created theoretical calibrations through the use of BasTi \citep{Basti_1,Basti_2} and STAREVOL \citep{starevol} respectively. These calibrations show disagreements between the theoretically predicted and observed level of \cn\ depletion during the FDU \citep{Cao_2025}. \citet{Casali_2019} and \citet{Spoo_2022} have created empirical calibrations with ages coming from open clusters \added{(Using Gaia-ESO \citep{GaiaESO_1,GaiaESO_2} with APOGEE DR14 \citep{dr14} or the OCCAMs catalog for APOGEE DR17 \citep{OCCAM}, respectively).} While these ages are trustworthy, the domain they can be used to calibrate is limited. \citet{Martig_2016} calibrated the relationship using the sample of field giants and asteroseismic masses and ages in APOKASC-2 \citep{apokasc2}.

As stated previously, the high interest in \cn-derived ages comes from the importance of ages for the study of the Milky Way. Reliable spectroscopic ages will significantly enhance the information that can be derived from large spectroscopic surveys, particularly in terms of the Milky Way's chemical history. The Milky Way shows variations in chemical abundance based on location, such as the radial metallicity gradient \added{\citep[e.g.][]{WilsonRood_1994, Anders_2017, Magrini_2023}} or the difference in $\alpha$-enrichment between the thin and thick disks \citep[e.g.][]{Bensby_2003, Bensby_2014, Hayden_2015}. However, to understand how the Milky Way evolved into what we see, we must examine how abundances vary with time, as well as position. By examining stars with known ages, we can learn about the gas and dust at the time of their birth. 

Unfortunately, dynamical interactions from many sources can cause stars to move great distances from their birth locations \citep{SellwoodBinney_2002, Quillen_2009, SchonrichBinney_2009}. Beyond even this, large bulk trends can be influenced by galaxy-wide events such as infalls of pristine gas and variations in star formation \added{\citep{Chiappini_1997, VICE, Spitoni_2023, Spitoni_2024, Palla_2024} Galactic mergers can drastically impact a galaxy's chemical profile \citep{TapiaContreras_2025}, which the MW has definitely experienced over its lifetime \citep{Ibata_1994, Belokurov_2018, Deason_2024}. Such effects must be considered when using using stars to learn about the history of our Galaxy.}

This work seeks to improve upon previous \cn-age calibrations with the larger samples of stars and more robust data pipelines now available. With the larger samples of galactic surveys, we can then apply our age calibrations to a wide range of field stars to examine aspects of the Galactic history. The paper is structured as follows. Section \ref{sec:data} describes our data sources and the selection criteria for our samples. Section \ref{sec:fitting} details the process to determine the \cn-Age relationship and provides the final equation. Section \ref{sec:validity} examines where the relationship holds and how accurate it is, and discusses the power of \cn\ as an age diagnostic. Section \ref{sec:archeology} explores applications of these ages to examine the chemical evolution of the Milky Way. Finally, Section \ref{sec:conclusions} summarizes our findings and discusses potential extensions to this work.
\section{Data and Sample Selection} \label{sec:data}

We use asteroseismic data on masses and ages to calibrate the spectroscopic age diagnostics. Here, we introduce the data that we used for this effort. Sections \ref{data:spectro} and \ref{data:APOKASC} describe the sources of our spectra and asteroseismic data, respectively. Section \ref{data:samples} describes the different samples we select for our analysis in this paper.

\subsection{Data Sources}\label{data:sources}
\subsubsection{Spectra}\label{data:spectro}

The Apache Point Observatory Galactic Evolution Experiment \citep[APOGEE;][]{Majewski2017} was part of the Sloan Digital Sky Survey (SDSS), in particular SDSS-III \citep{eisenstein2011} and SDSS-IV \citep{blanton2017}. It collected high-resolution H-band spectra using dual APOGEE spectrographs \citep{Wilson2019} at the 2.5-meter Sloan Foundation Telescope \citep{Gunn2006} at Apache Point Observatory and the 2.5-meter Ir{\'e}n{\'e}e DuPont telescope \citep{Bowen1973} at Las Campanas Observatory. Abundances, effective temperatures, and surface gravities used in this paper come from the 17th data release of SDSS \citep[][; DR17]{dr17}. The spectra were reduced by the APOGEE data reduction pipeline \citep{Nidever2015}. The stellar parameters and abundances were determined by the APOGEE Stellar Parameter and Chemical Abundances Pipeline \citep[ASPCAP;][]{aspcap}, which compares the observed spectra with a large grid of synthetic spectra. The effective temperatures, surface gravities, and abundances are then placed on an absolute scale in a post-processing calibration step, and cases with suspect or bad overall fits are flagged. A description of the APOGEE flags can be found in \citet{jonsson2020}.

APOGEE DR17 has a large number of Value Added Catalogs (VACs) created by SDSS collaborators, which provide additional parameters for the stars in APOGEE DR17. We combine the spectroscopic abundances with distance and position parameters derived in the AstroNN VAC \citep{AstroNN_1, AstroNN_2, AstroNN_3}, 
\added{based on input Gaia data \citep{Gaia_EDR3}. In this paper, we use the AstroNN} Galactocentric X, Y, Z, and R coordinates, as well as the guiding radius and maximum Z height parameters.

We adjusted the spectroscopic abundance errors from the DR17 release catalog based on \citet{apokasc3}. We assume a minimum uncertainty of 0.05 in \feh\ for all stars, following their methods. Additionally, for [C/Fe] and [N/Fe] uncertainties, we adopt the changes prescribed by \citet{Cao_2025} in their appendix, increasing reported errors by factors of 3.0278 and 2.7109, respectively. 

\subsubsection{Asteroseismology} \label{data:APOKASC}

Masses, ages, and evolutionary states in this paper come from the APOKASC-3 and APO-K2 catalogs. 

APOKASC-3 includes stars targeted by the original \kep\ mission \citep{Kepler} and the APOGEE survey. There have been 3 catalogs in total \citep{apokasc1, apokasc2, apokasc3}, with a total of 15,808 evolved giants in the third complete data set. Stellar parameters from seismology are calculated through 10 independent pipelines and compared to ensure accuracy, and the average of those pipelines is then added to the catalog. A more complete description of the processes can be found in \citet{apokasc3}. Asteroseismology can be used to infer evolutionary states and to calibrate spectroscopic evolutionary state predictions. For a discussion on how these predictions are made, see \citet{Elsworth_2019, Warfield_2021}, and \citet{apokasc3}. With an evolutionary state assigned, the stellar parameters are used to query an interpolated set of standard stellar models to find an associated age. A full description of this process and the models can be found in \citet{apokasc3}. 

The APO-K2 catalog \citep{Stasik_2024, Warfield_2024} utilizes a similar analysis to that employed in the APOKASC-3 catalog for targets from the K2 survey \citep{K2}. It contains 7672 red giants with known masses and evolutionary states. 4661 of the first-ascent giants have age determinations. As these stars have shorter baselines of observation, the uncertainties are much larger than for the APOKASC-3 sample. However, the K2 survey samples a much wider range of populations from many different points in the Milky Way. Therefore, we can use it to ensure that our data, though calibrated exclusively on the \kep~field, is representative of stars all around the Galaxy.

\subsubsection{Cluster Membership and Age Parameters}

Ages can be inferred for open clusters independent of the method used for asteroseismology. Open cluster members, therefore, provide an external check on our asteroseismic age scale. We reference the Open Cluster Chemical Analysis and Mapping (OCCAM) catalog \citep{OCCAM}, which uses data from both APOGEE and Gaia DR3 to establish membership probabilities for APOGEE stars in open cluster fields.
We use the membership probabilities from the OCCAM catalog and take ages derived by \citet{Cantat-Gaudin_2020}, who used an artificial neural network trained on reference clusters to predict cluster ages and distances for clusters observed in Gaia DR2. \added{We give all clusters an assumed uncertainty of 0.1 in log(age) as done in \citet{Spoo_2022}}. By combining these data sets, we can create a sample of clusters with known ages and members with APOGEE abundances.

\subsection{Sample Selection}\label{data:samples}

We separate the data used in this paper into three different categories. First, there is the calibration sample, taken from APOKASC-3, which we use to determine the [C/N]-age relationship used throughout the paper. Second, there is the validation sample, which consists of stars taken from APO-K2 and red giants in open clusters. We use the validation sample to ensure that the relationship we determine, though calibrated on \kep~field stars, can be applied to a larger sample of stars. Finally, we have the mapping sample, comprised of red giants from APOGEE DR17. Our galactic archaeology results are found by applying the [C/N]-age relationship to the mapping sample to obtain ages for stars throughout the galaxy.  

\subsubsection{Evolutionary State Distinctions}\label{data:evstate}

All three samples include stars of different evolutionary states. Differences in physical processes and population distributions between these states necessitate the ability to distinguish them from each other. In this paper, we separate stars into four different evolutionary states: Lower Red Giant Branch (LRGB) stars, Upper Red Giant Branch (URGB) stars, Red Clump (RC) stars, and Asymptotic Giant Branch (AGB) stars. LRGB and URGB stars are first-ascent giants that we separate at the RGB bump. The bump is an over-density of stars on the RGB caused by temporary re-contraction during RGB evolution. The H-burning shell reaches the chemical discontinuity left by the convective envelope during the FDU and temporarily slows energy generation, causing the stars to ''pile up`` on the RGB around \logg\ of 2.5. The RC stars have completed their first ascent and have re-contracted to begin core-He burning. AGB stars have finished core-He burning and have begun their second ascent while shell-He burning. 

We separate the stars into these categories to handle concerns of extra mixing, mass loss, and the reliability of asteroseismic parameters for very luminous giants. In some age and metallicity domains, there is strong observational evidence for extra mixing in luminous giants \citep[e.g.][]{Kraft_1994}. The most likely explanation is thermohaline mixing \citep{Eggleton_2006, Charbonnel&Lagarde_2010}, caused by a small mean molecular weight inversion arising from non-equilibrium burning in the pp-chain. This mixing will continue to lower \cn\ and contaminate the relationship between \cn\ and age in the stars where it occurs. Extra mixing affects the surface abundances for stars of \feh$<$-0.4 \citep{Shetrone_2019} and so \cn\ is only a useful age diagnostic for LRGB metal-poor stars at present. Additionally, stars experience mass loss on the RGB, primarily at the tip of the RGB just before and during the transition to the RC phase \citep{Reimers_1975}. Asteroseismology measures the current mass, whereas \cn\ correlates with mass during the FDU. In APOKASC-3, mass loss was calibrated to require consistent ages for the alpha-rich RGB and RC stars, so the age assignments in that catalog do account for mass loss. However, the empirical mass loss expression used in those models is not rigorous or strongly motivated by theory. As a consequence, we separate the RC and AGB stars from the LRGB and URGB to test the consistency of the results for the alpha-poor population.
The final critical distinction between these states deals with luminosity. Asteroseismic scaling relations break down as stars become more luminous \citep{Ash_2025}. In addition, separating AGB from URGB stars is difficult to do as the two branches have very similar temperatures. Separating the URGB and AGB from the LRGB and RC allows us to separate the stars with more confident asteroseismic ages from the more uncertain ones.

In APOKASC-3, RGB, RC, and AGB stars have been separated asteroseismically. We separate the RGB into LRGB and URGB at asteroseismic \logg=2.5, the highest extent of the RGB bump in the sample. For stars coming from APO-K2, only first ascent giants have calculated ages, and so only those are used. The LRGB and URGB distinction remains the same as for the APOKASC-3 stars.

For stars coming solely from APOGEE DR17, no asteroseismic evolutionary state determination is available. Instead, we use the RGB/RC separation criteria employed in the full APO-K2 catalog \citep{Warfield_2024, Stasik_2024}. The criteria compares a star's effective temperature against the expected temperature for an RGB star at that gravity and metallicity. Stars hotter than a certain threshold, calibrated using APOKASC-3 data, are flagged as RC stars. As before, the same \logg\ line was used to separate LRGB and URGB.

\subsubsection{Calibration Sample}\label{data:calib}

\begin{figure}[ht!]
\includegraphics[width=\columnwidth]{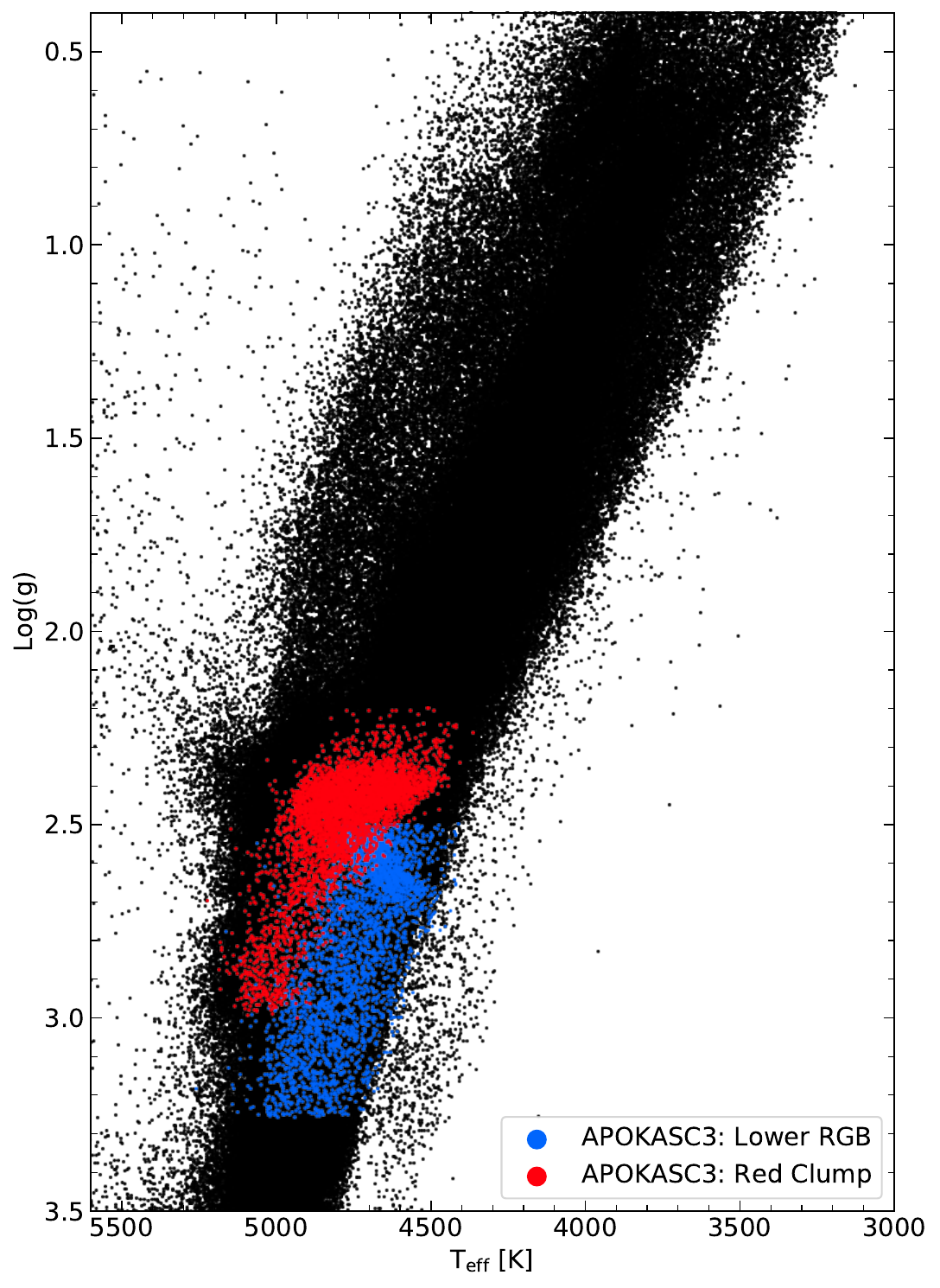}
\caption{Kiel diagram of the APOKASC-3 stars used in our calibration sample. The black points in the background are the full APOGEE DR17 sample, provided for context. The blue stars are LRGB stars, and the red stars are RC stars, determined asteroseismically. 
\label{fig:kascHR}}
\end{figure}

Our calibration sample consists exclusively of stars taken from APOKASC-3, with some cuts to ensure high-quality data. First, we limited our sample to the 10,136 star ``gold sample'' flagged in the catalog, which are the stars with the lowest uncertainties and most reliable parameters. Additionally, stars with asteroseismic \logg$\geq3.26$ were removed to ensure that the sample has completed the FDU \citep{Roberts_2024}. We do not use the more luminous giants (URGB and AGB) for calibration, due to their asteroseismic scaling concerns. To restrict contamination from extra mixing, RC stars with \feh\ below -0.4 have been omitted. We remove stars with \cn\ below -0.75 as such stars fall well outside the range of \cn\ spanned by the majority of the sample and are likely results of non-standard evolution. Finally, there are still two stars with abnormally high nitrogen uncertainties (in excess of $\pm$ 2 dex) that we remove. The total list of criteria we apply to the APOKASC-3 dataset to create our calibrating sample is given below. The stars remaining in the calibrating sample are shown in Figure \ref{fig:kascHR}.

\begin{equation}
    \begin{cases}
      \text{Catalog: Gold} \\
      2.5 \leq \mathrm{\logg} \leq 3.26~\mathrm{OR ~ Evol\_State=RC} \\ 
      \mathrm{\cn} \geq -0.75\,~. \\
      \mathrm{Age (Gyr)} \leq 13.2 \\
      \sigma\mathrm{\nfe} < 2 \\
      \text{If RC:} \\
      -0.4 \leq \mathrm{\feh}
    \end{cases}       
\end{equation}

\subsubsection{Validation Sample}\label{data:valid}

The validation sample has two components: red giants that are members of open clusters and those in APO-K2. 

The clusters used in this paper are the high-quality clusters older than 100 Myr as defined in \citet{OCCAM}. We included only stars with membership probabilities of 70\% or higher and stars that were not tagged as potential members of more than one cluster.

\begin{figure}[ht!]
\includegraphics[width=\columnwidth]{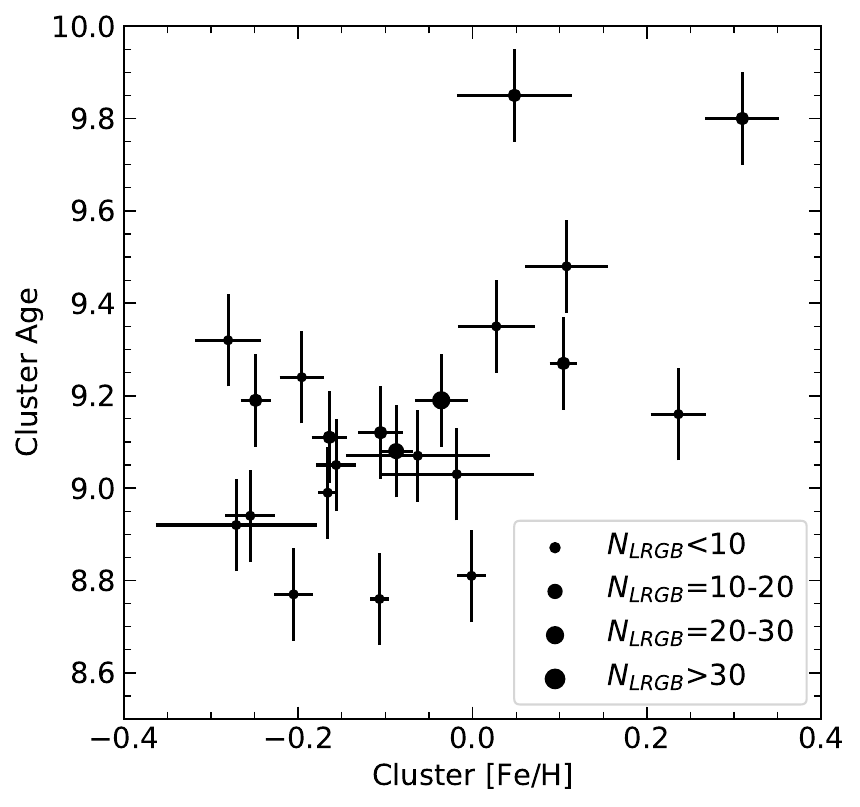}
\caption{Age-\feh\ for the 21 clusters used in this paper. 
\label{fig:clustersample}}
\end{figure}

We also take advantage of the internal chemical homogeneity of open clusters in \feh to remove possible non-members. We removed candidate members with \feh\ greater than 1.5 sigma from the mean of the cluster. The mean \cn\ was calculated from the LRGB stars in the sample, defined as 2.5 $<$ \logg\ $<$ 3.26 and 4400 $<$ \teff\ $<$ 5300 K. If any cluster had fewer than 3 stars meeting these criteria, it was removed from consideration. This leaves 21 open clusters for comparison, shown in Figure \ref{fig:clustersample}.

For the APO-K2 validation sample, we used only the LRGB stars because RC/AGB stars were not assigned ages in APO-K2, and we do not use URGB stars for calibration. The boundary criteria used for the APO-K2 stars are given below:

\begin{equation}
    \begin{cases}
      \mathrm{\cn} \geq -0.75\,~. \\
      \mathrm{Age (Gyr)} \leq 13.2 \\
      2.5 \leq \mathrm{\logg} \leq 3.26
    \end{cases}       
\end{equation}

\begin{figure}[ht!]
\includegraphics[width=\columnwidth]{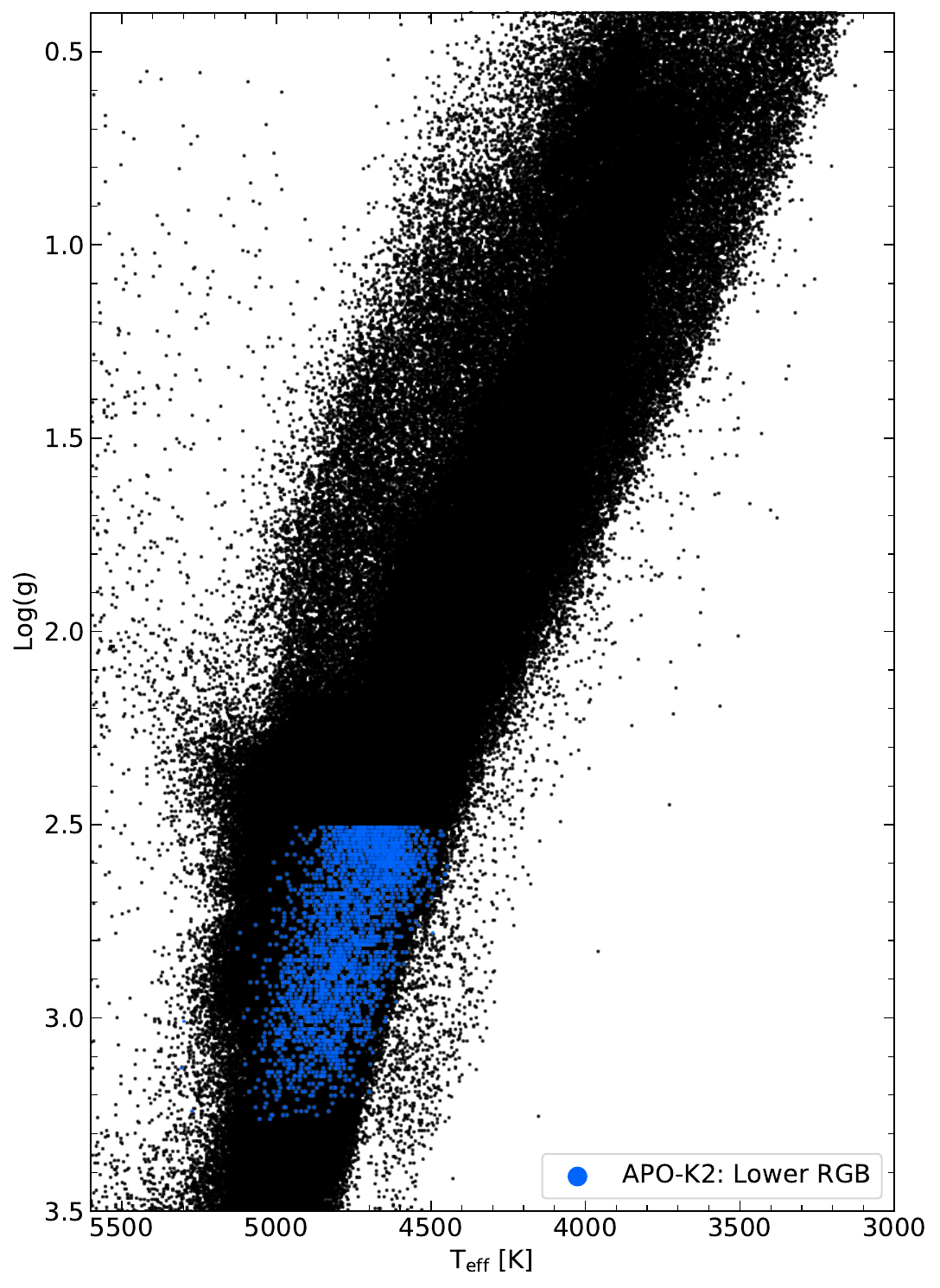}
\caption{Kiel diagram of the APO-K2 stars used in this paper. The conventions and labeling are identical to Figure \ref{fig:kascHR}.}
\label{fig:K2HR}
\end{figure}

\subsubsection{Mapping Sample}

The mapping sample is comprised of red giants from APOGEE DR17.
\added{We consider stars with \logg\ values between 3.26 (to ensure stars have completed the FDU) and 0.5 (to avoid APOGEE issues with low gravity stellar parameters). We removed stars with \cn\ below -0.75 as we did for the calibration sample. We also removed stars with flags indicating complex history or unusual parameters: possible young cluster member, emission line star, MIR-detected candidate cluster member, part of the eclipsing binary program, or part of a W3/4/5 star-forming complex. These stars do not have asteroseismic states, so we use the same spectroscopic criteria that we used for the APO-K2 sample \citep{Stasik_2024, Warfield_2024}. In this process, a characteristic temperature for an RGB star is calculated and compared with the star's measured temperature. Our final cut removed  5$\sigma$ outliers in this $\Delta T$ domain, determined separately for RGB and RC stars.} The full list of criteria for the mapping sample can be found below:

\begin{equation}\label{eq:mappingcrit}
    \begin{cases}
      0.5 \leq \mathrm{\logg} \leq 3.26 \\ 
      \mathrm{\cn} \geq -0.75\,~. \\
      \sigma\mathrm{\cn} \leq 0.1 \\
      \text{If RGB:} \\
      -515 \leq T_\mathrm{Ref}-T_\mathrm{eff} (\mathrm{K}) \leq 340 \\
      -0.4 \leq \mathrm{\feh}~(\mathrm{If}~\logg<2.5) \\
      \text{If RC:} \\
      -0.4 \leq \mathrm{\feh} \\
      -620 \leq T_\mathrm{Ref}-T_\mathrm{eff} (\mathrm{K}) \leq 100 \\
    \end{cases}       
\end{equation}

\begin{figure}[ht!]
\includegraphics[width=\columnwidth]{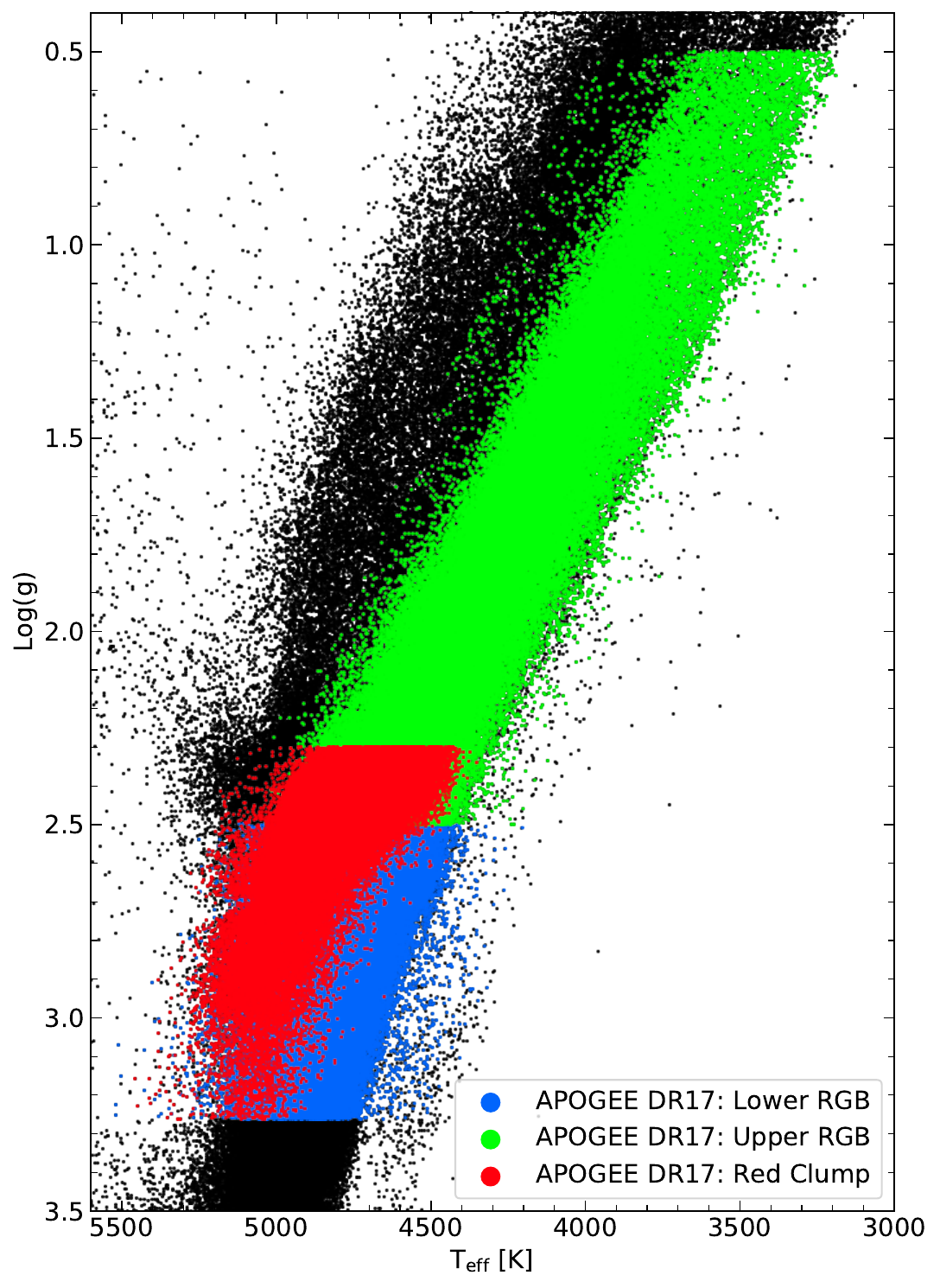}
\caption{Kiel diagram of the APOGEE DR17 stars used in this paper. All conventions remain the same as Figure \ref{fig:kascHR}. The stars on the left (hot) side of the giant branch are excluded due to our extra-mixing cut, which removes the metal-poor stars. 
\label{fig:geeHR}}
\end{figure}

\begin{figure*}[ht!]
\includegraphics[width=2.1\columnwidth]{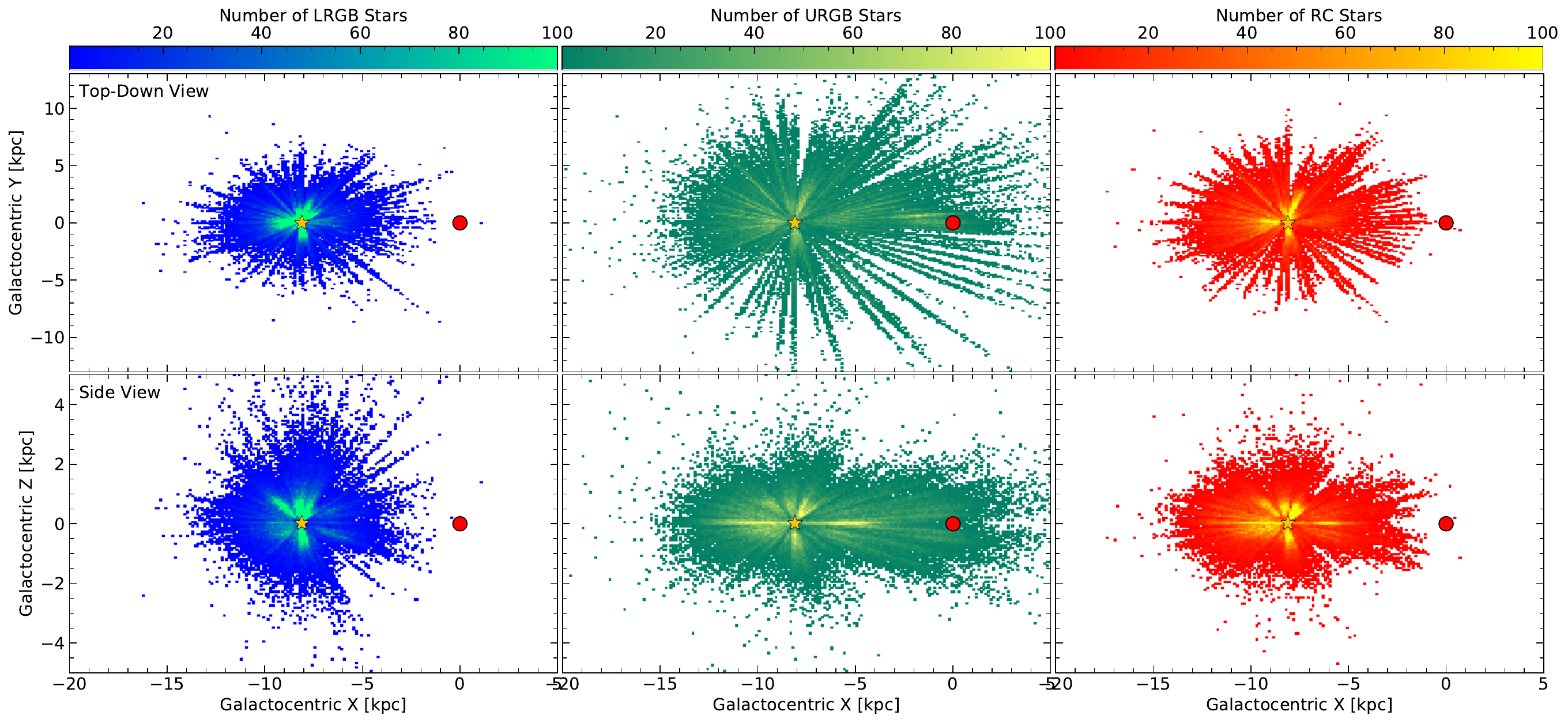}
\caption{Galactocentric X-Z map of the APOGEE DR17 giants used in this paper, separated by evolutionary state. The yellow star marks the Sun, and the red circle marks the galactic center.}
\label{fig:geemapevstate}
\end{figure*}

The stars in the mapping sample are shown in Figure \ref{fig:geeHR}. \added{Figure \ref{fig:geemapevstate} shows the distributions of the mapping sample in galactic coordinates, with each evolutionary state shown individually}. Predictably, the LRGB and RC stars occupy similar scales, and the URGB stars, being more luminous, extend to farther distances. It is therefore valuable for Galactic archaeology to derive a robust age diagnostic for these stars.

\section{The Age-\cn\ Relationship}\label{sec:fitting}

With our sample of stars with known ages, we can calibrate a relationship to determine stellar age from surface chemistry. In Section \ref{fit:method}, we discuss the process used to obtain our fits. In Section \ref{fit:LBRC}, we obtain fits for the LRGB and RC stars separately to determine if there are systematics between the two evolutionary states. Finally, in Section \ref{fit:cnfit} we present the calibrated \cn-Age relationship.

\subsection{Fitting Methods}\label{fit:method} 

Our process for obtaining a fit to the relationship between \cn, \feh, and age has two primary steps, similar to those used in \citet{Roberts_2024}. First, we bin the points across both \cn\ and \feh\ and divide them into 20 cohorts of equal size in both \feh\ and \cn. Then, each bin was assigned a value of \feh, \cn, and the age of the error-weighted median of all stars in the bin.

Next, a polynomial function of the form in Equation \ref{eq:fit} was fit to the data. Higher-order polynomials and non-polynomial functional forms were tested, but did not result in more accurate ages.

    \begin{equation}\label{eq:fit}
    \begin{aligned}
        \text{Log(Age/year)} = & c_5(\cn)^2 + c_4(\cn) + \\
        &c_3(\text{[Fe/H]})^2 + c_2(\text{[Fe/H]}) + \\
        &c_1(\cn)(\text{[Fe/H]}) + c_0^{}\,~,
    \end{aligned}
    \end{equation}

\subsection{Age consistency between the LRGB and RC}\label{fit:LBRC}

In the absence of extra mixing, the \cn\ set by the FDU should not change until the tip of the AGB and the third dredge-up \added{\citep{Karakas_2014}}. We have omitted metal-poor stars where extra mixing is known to be an issue \added{\citep{Shetrone_2019}}. Therefore, the Age-\cn\ calibration should be the same for all of the stars in our sample. 

The RC can sample younger stars than the LRGB because stars that rapidly evolve through the first ascent remain in the RC longer, increasing the sample size. By combining the two, we can extend our calibration to wider ranges of ages than could be reliably done with a single sample alone. 

\begin{figure*}
    \includegraphics[width=1.99\columnwidth]{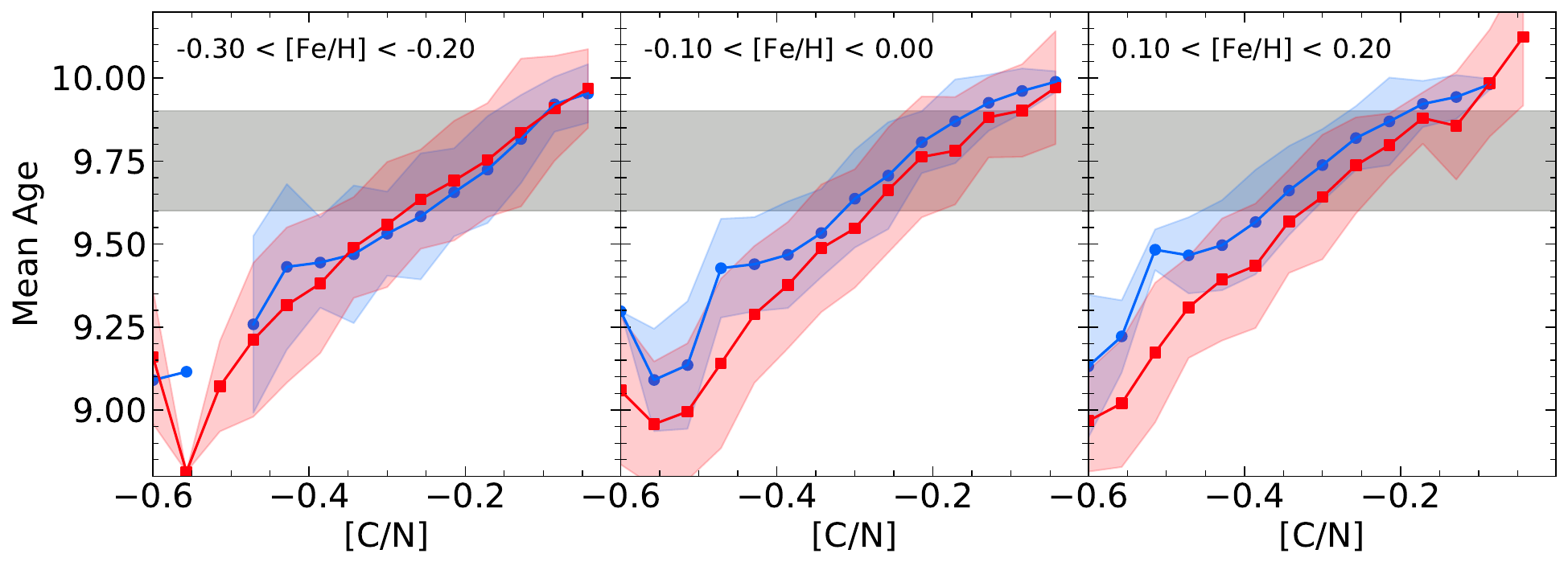}
    \caption{Age vs \cn\ in different windows of \feh. The points reflect the mean values of a bin of stars about that \cn. The shaded region shows the 1-$\sigma$ bounds. Blue points are LRGB stars, and red points are RC stars.}
    \label{fig:RCadjustment}
\end{figure*}

\added{As mentioned in the introduction, asteroseismic ages are dependent on the treatment of mass loss assumed in the models. Given the sensitivity of stellar age to mass, even slight inaccuracies in mass loss prescriptions could cause notable issues with ages in Red Clump stars, as compared to first ascent giants.} In APOKASC-3, the mass loss was adjusted such that the ages of alpha-rich RC and LRGB stars were consistent. However, this was not checked in the alpha-poor sample. It is also possible that there are temperature and metallicity-dependent offsets between the RGB and RC in the derived \cn. \added{To ensure that the LRGB and RC samples are in proper agreement,} we compare the two states separately before merging them. Figure \ref{fig:RCadjustment} shows the difference between the LRGB and RC for three different metallicities. The two trends are quite close overall. In the lower metallicity panel, the differences are negligible, but as metallicity increases, RC stars are assigned younger ages than the LRGB counterparts at comparable \cn\ and \feh.

We apply a correction to the RC ages to bring them in line with the LRGB ages, since the cause of this difference is mass loss. The LRGB reflects the correct, unaltered relationship. We obtain our correction by fitting the LRGB and RC independently using the method described in Section \ref{fit:method} with a smaller number of bins and cohorts to keep a similar number of stars per bin. Then, for a given metallicity, we determined the range in \cn\ that the LRGB spanned over an age range of 9.6-9.9 dex in log(age). For this \cn\ range, the median offset between the ages of two fits was tracked, and we fit a polynomial function to create our correction function below:

\begin{equation}
    \Delta\mathrm{log(Age)}_{RC} = -0.498\times\mathrm{\feh}^2 + 0.253\times\mathrm{\feh} + 0.079
\end{equation}

Over the \feh\ range of our sample, this correction adjusts the ages of RC stars by between -0.09 and +0.09 dex. The form of this correction implies that the metallicity dependence of the mass loss prescription used in the APOKASC-3 model grid is slightly overestimated. \added{This correction is applied to the stars before we perform our final calibration, described in the following section.}

\subsection{The [C/N] - Age Relationship(s)}\label{fit:cnfit}

\begin{figure*}
    \includegraphics[width=2.2\columnwidth]{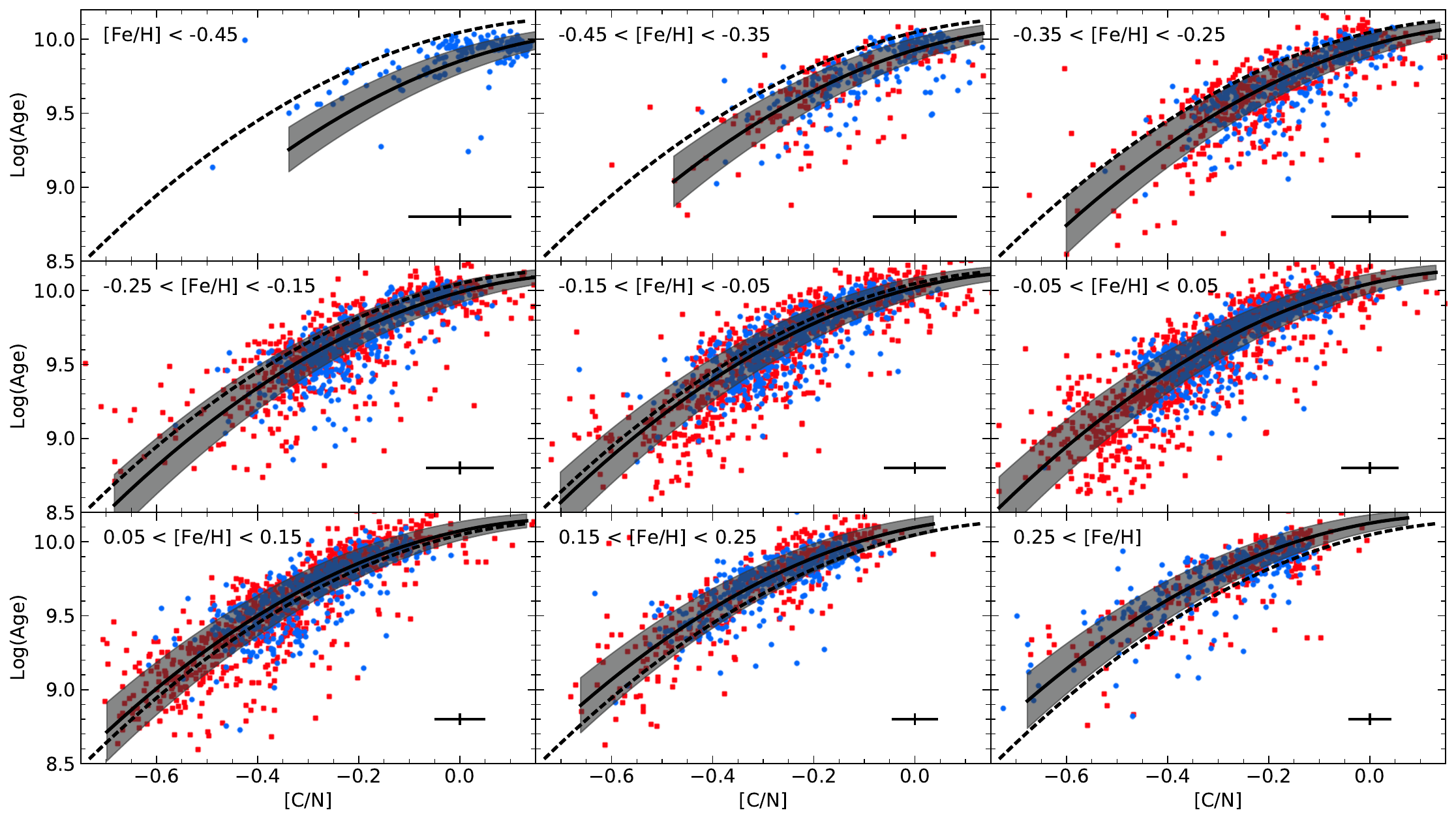}
    \caption{Age vs \cn\ in different windows of \feh. Blue points are LRGB stars, and red points are RC stars. The error bars at the bottom right of each panel show the median uncertainty values of the points in the bin. The \feh=0 trend is shown in all panels as a dashed line, for comparison. The shaded region is the uncertainty of the fit derived from the typical observational uncertainties.}
    \label{fig:primarypanel}
\end{figure*}

With the RC and LRGB aligned, we use the method described in Section \ref{fit:method} to create our \cn-Age relationship. Table \ref{tab:fitprimary} has the coefficients for this function. Figure \ref{fig:primarypanel} shows the function overlaid on the data. 

\begin{table}
    \centering
    \begin{tabular}{c c c c c c c}
         \cn\ & $c_5$ & $c_4$ & $c_3$ & $c_2$ & $c_1$ & $c_0$  \\
        \hline      
         = & -1.721 & 0.806 & -0.077 & 0.276 & -0.6430 & 10.048 \\
    \hline
    \end{tabular}
    \caption{\cn-Age Function Coefficients}
    \label{tab:fitprimary}
\end{table}
    
\section{Applicability of [C/N] as an Age Diagnostic}\label{sec:validity}

We have the ability to predict age for some red giants of a given \cn\ and \feh, but we must understand where this relationship can be used, and how accurate it is. In Section \ref{val:recover}, we examine how well our function recovers the training ages and set bounds for its application. In Section \ref{val:limits}, we apply our fit to our validation samples and compare with previous empirical calibrations. In Section \ref{val:luminous}, we show how the function can be applied to luminous giants despite their exclusion from the training sample. Finally, in Section \ref{val:theory}, we compare our results with previous theoretical calibrations.

\subsection{Recovery Tests}\label{val:recover}

\begin{figure}
    \includegraphics[width=\columnwidth]{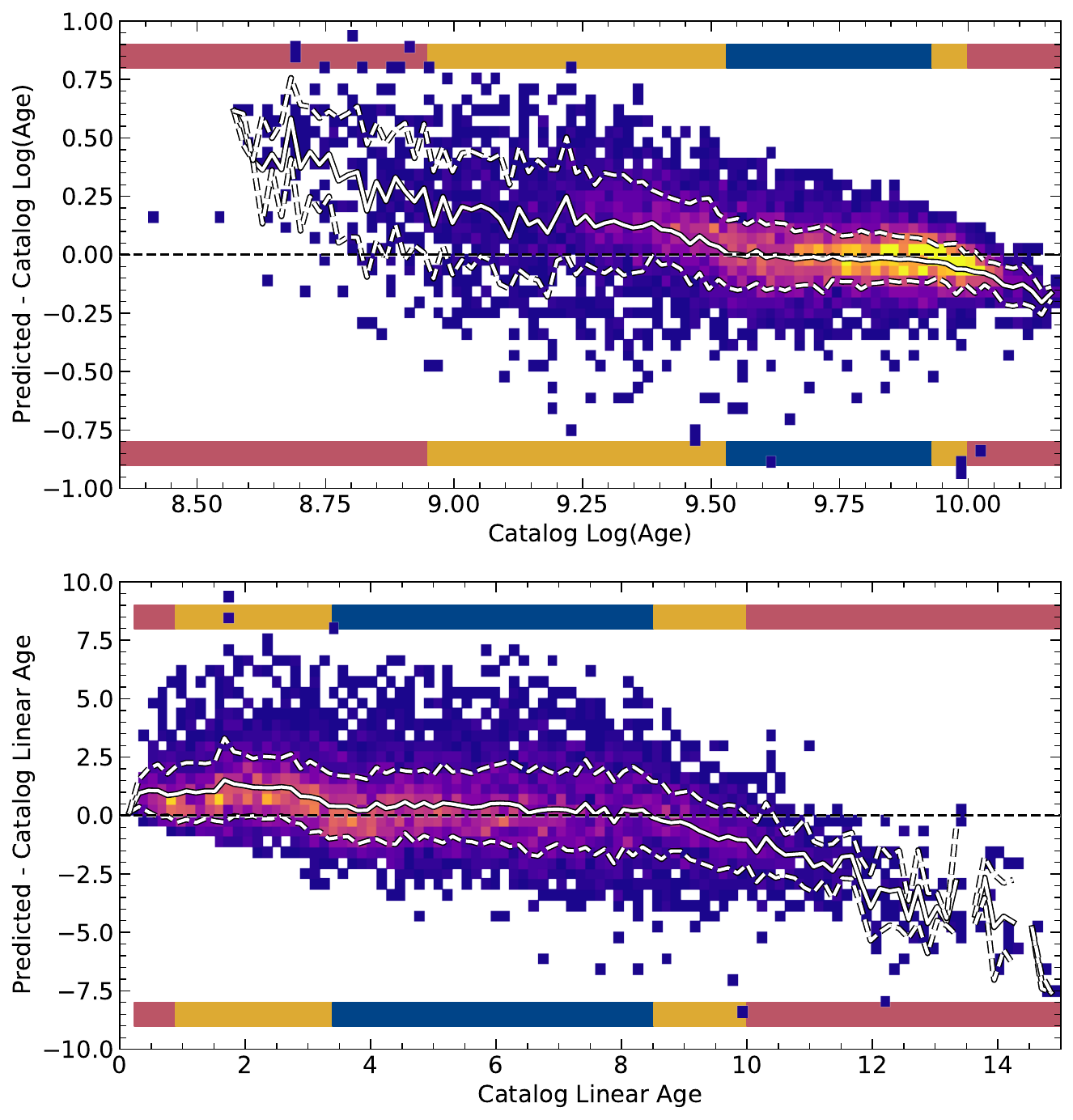}
    \caption{Difference between our \cn\ predicted age and the catalog age for the calibration sample. High values indicate the relationship over-predicting age and vice versa. The white solid line shows the mean of the residuals at the corresponding age, and the dashed lines show the 1-$\sigma$ bounds. Blue, yellow, and red bars show the optimal, fair, and poor regions, respectively.}
    \label{fig:ageresiduals}
\end{figure}

Figure \ref{fig:ageresiduals} shows the residuals between the catalog ages and our predicted ages in both linear and logarithmic space, with lines showing the mean and standard deviation of the residuals at each point. The bulk of the distribution has evenly distributed residuals, but at the youngest and oldest edges of the domain, the fit fails. The scatter is also greater than would be expected from observational uncertainties alone. The RMS deviation of log(age) of all stars is 0.174, whereas propagating the observational uncertainties through our relationship results in a median uncertainty of 0.114. This excess scatter is likely due to variations in birth carbon and nitrogen for stars of a given \feh. For this reason, we opt to use the residual scatter as our indication of uncertainty.

\added{The RMS scatter varies with age, being growing in the domains near where the fit fails. Interestingly, however, the RMS scatter in linear space is nearly constant across most of the domain. For this reason, as well as the preferred use of linear ages in Galactic archaeology, we transition to discussing the fit in the linear domain, rather than log.} 

Using the RMS scatter, we divide the domain into three regions: optimal, fair, and poor. The optimal region \big(3.34 Gyr $\leq$ Age $\leq$ 8.5 Gyr\big) has no strong bias in the residuals. This is the region where \cn\  can be reliably used as an age indicator with our calibration. The fair region \big(0.9 Gyr $\leq$ Age $\leq$ 10 Gyr\big) has a slight bias in the residuals but still less than the scatter of the distribution. This was defined as having the mean of the residuals differ from 0 by more than 0.3, but less than $1\sigma$. This is the domain where \cn\ has some predictive power, but caution should be used when interpreting results. Finally, the poor region \big(Age $\leq 0.9$ Gyr or $\geq 10$ Gyr\big) contains residuals that are very strongly biased. This is the region where \cn\ cannot be reliably used as an age indicator. Within the optimal and fair regions, the RMS scatter has a nearly constant mean value of 1.64 Gyr. Although we calibrated our ages in logarithmic space, we recommend using 1.64 Gyr as the uncertainty, as this value remains constant across both the optimal and fair domains.

The locations and boundaries of these domains make physical sense. For the older stars, the mass differences between ages become small enough that \cn\ changes between stars are too small to reliably distinguish between age \citep{Mackereth_2019}. Conversely, the \cn\ for younger stars loses predictive power due to the physics of the FDU. As shown in \citet{Roberts_2024}, as mass increases, \cn\ becomes insensitive to mass above 2.2 solar masses, which corresponds to $\sim$1 Gyr. This effect can be seen in other \cn-age calibrations as well, such as \citet{Martig_2016} and \citet{Stone-Martinez_2024}.

\subsection{Validation and Reliability Tests}\label{val:limits}

Our fit was trained on high-quality, precise data, but only for stars within the \kep\ field and only on a single self-consistent system of ages. \added{We now test our fits on our validation sample to show they can be applied to the broader Galaxy.}

\subsubsection{Validity outside the \kep~Field}

\begin{figure}
    \includegraphics[width=\columnwidth]{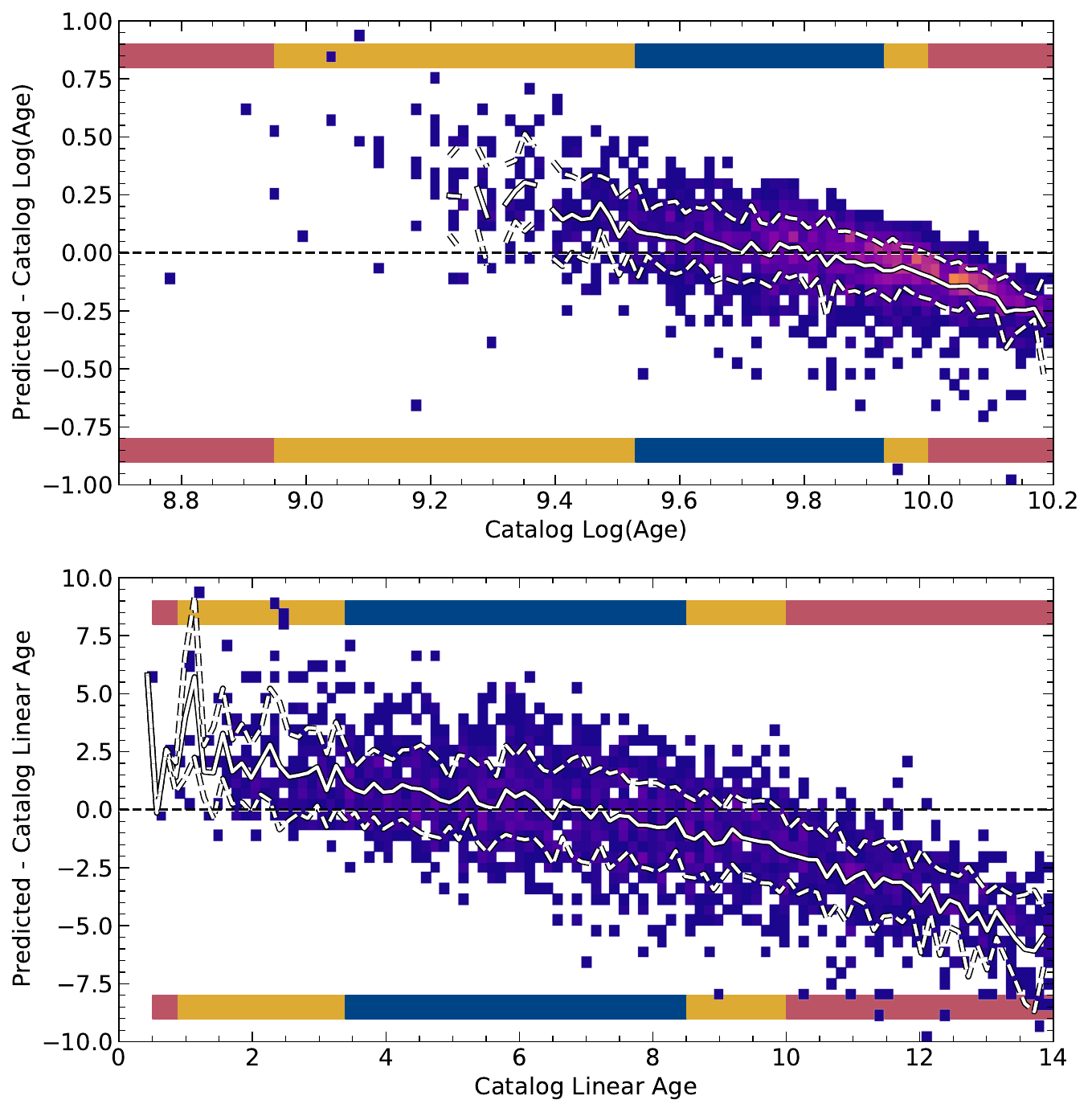}
    \caption{Difference between our \cn\ predicted age and the catalog age for the K2 sample. The axes and conventions are the same as for Figure \ref{fig:ageresiduals}. The uncertainty of the K2 ages is higher than our calibration sample, which shows in the worse agreement in the fair and poor regions. However, the regions still accurately reflect the behavior of the data.}
    \label{fig:K2check}
\end{figure}

Figure \ref{fig:K2check} shows our fit compared to the APO-K2 stars. The RMS scatter is higher for this sample, at about 2.18 Gyr within the optimal and fair regions (once again, more consistent in linear space). However, the median propagated uncertainty within this region is 2.05 Gyr, closer to the scatter than for the calibration sample. When considering the increased uncertainties, our fit performs similarly on the APO-K2 sample as it does for the APOKASC-3 sample. This means that our fit is valid across different regions of the galaxy, though only calibrated on stars in the \kep\ field.

\subsubsection{Comparisons with non-asteroseismic ages}

\begin{figure}
    \includegraphics[width=0.9\columnwidth]{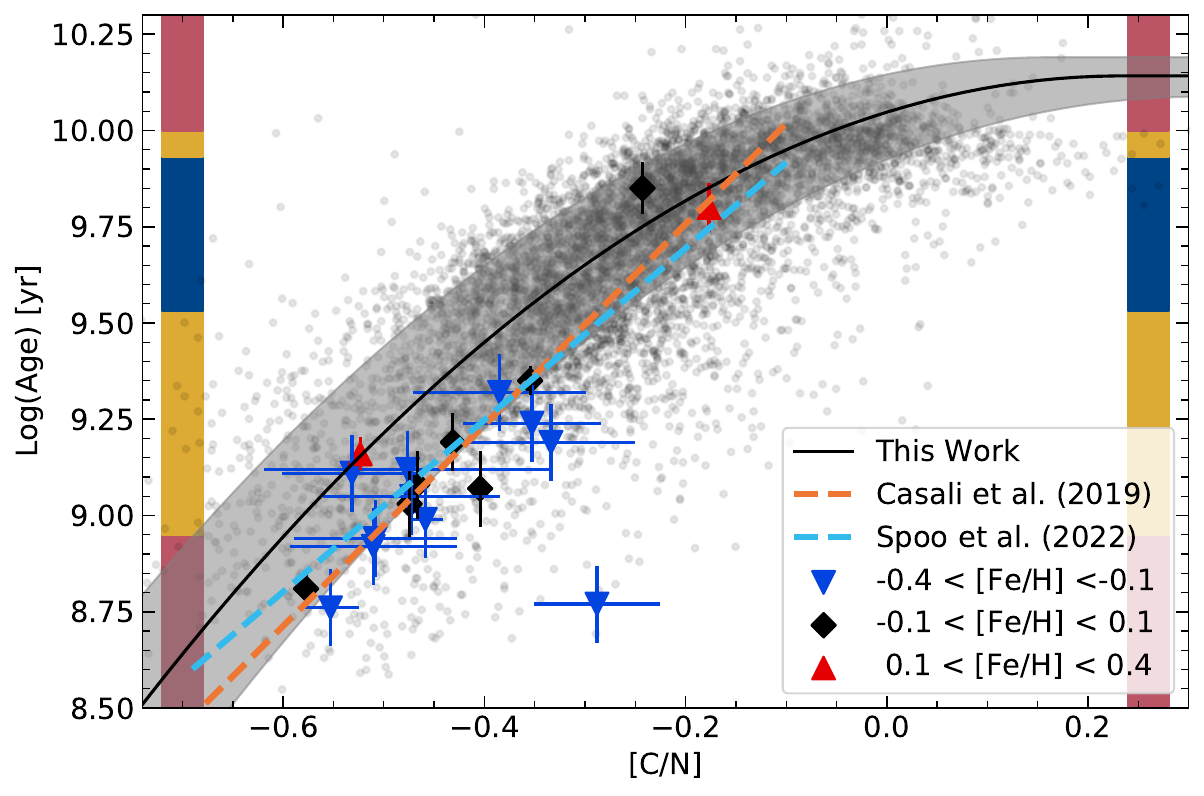}
    \caption{Expected age for a given \cn\ using fits from this work, \cite{Spoo_2022}, and \cite{Casali_2019}. The line represents the solar metallicity fit, and the shaded region spans differences of 0.4 dex in either direction from solar. Each diamond represents a cluster of known age and RGB \cn, colored by three bins of \feh. The transparent gray points are the calibration sample. The blue, yellow, and red bars on the side represent the optimal, fair, and poor regions, respectively, as in Figure \ref{fig:ageresiduals}}
    \label{fig:empcompare}
\end{figure}

Previous works have used clusters to calibrate the \cn-age relationship. \citet{Spoo_2022} calibrated using clusters in the OCCAMs catalog \citep{OCCAM} with abundances from APOGEE DR17. \citet{Casali_2019} used clusters of known ages from both the Gaia-ESO survey \citep{GaiaESO_1,GaiaESO_2} and the APOGEE DR14 survey \citep{dr14} to calibrate the relationship. We compare our fits with these published fits as well as the cluster RGB stars defined in Section \ref{data:valid}. Figure \ref{fig:empcompare} shows the fit compared to the cluster data and the fits from \citet{Casali_2019} and \citet{Spoo_2022}. As data reduction and analysis pipelines evolve, different APOGEE data releases may exhibit systematic offsets in abundance measurements for the same stars. To ensure these offsets do not influence our comparisons, we adjust the \cn\ used in plotting the fits from \citet{Casali_2019} 
by the offsets observed between the stars present in both data DR14 and DR17. This amounts to a shift down of 0.1 dex in the \cn. No \feh\ corrections were needed as their fit did not depend on \feh.

There is one cluster that matches none of the shown fits and is an outlier in regard to the rest of the clusters: Berkeley 2. Its \cn\ is unusually high in the LRGB stars, but if we consider all the giants, the disagreement becomes much smaller and within the range of the errors. For simplicity, we do not consider it for the calculations discussed below.

The metallicity independence of the two fits presents an interesting comparison. Our fit agrees well at \feh=-0.3, but diverges at higher metallicities due to the metallicity dependence we include. The clusters do show weaker metallicity dependence; however, the RMS deviation for the clusters is 0.177, practically identical to that for our calibration sample. Though we diverge at higher metallicities, our result is statistically consistent with ages derived from clusters.

\subsection{Applicability to Luminous Giants}\label{val:luminous}

We are now in a position to test the use of \cn\ as an age diagnostic for luminous giants. We examine the reliability of \cn\ for luminous giant ages by looking at red giants in open clusters NGC 6819, NGC 6791, and M67. The data for NGC 6819 and 6791 come from APOKASC-3, using the criteria from \citet{Ash_2025}. M67 abundance data were taken from APOGEE DR17 and combined with parallax and radial velocity data from Gaia EDR3 \citep{Gaia_EDR3} and the selection criteria used in \citet{Reyes_2024}.

\begin{figure*}
    \includegraphics[width=1.95\columnwidth]{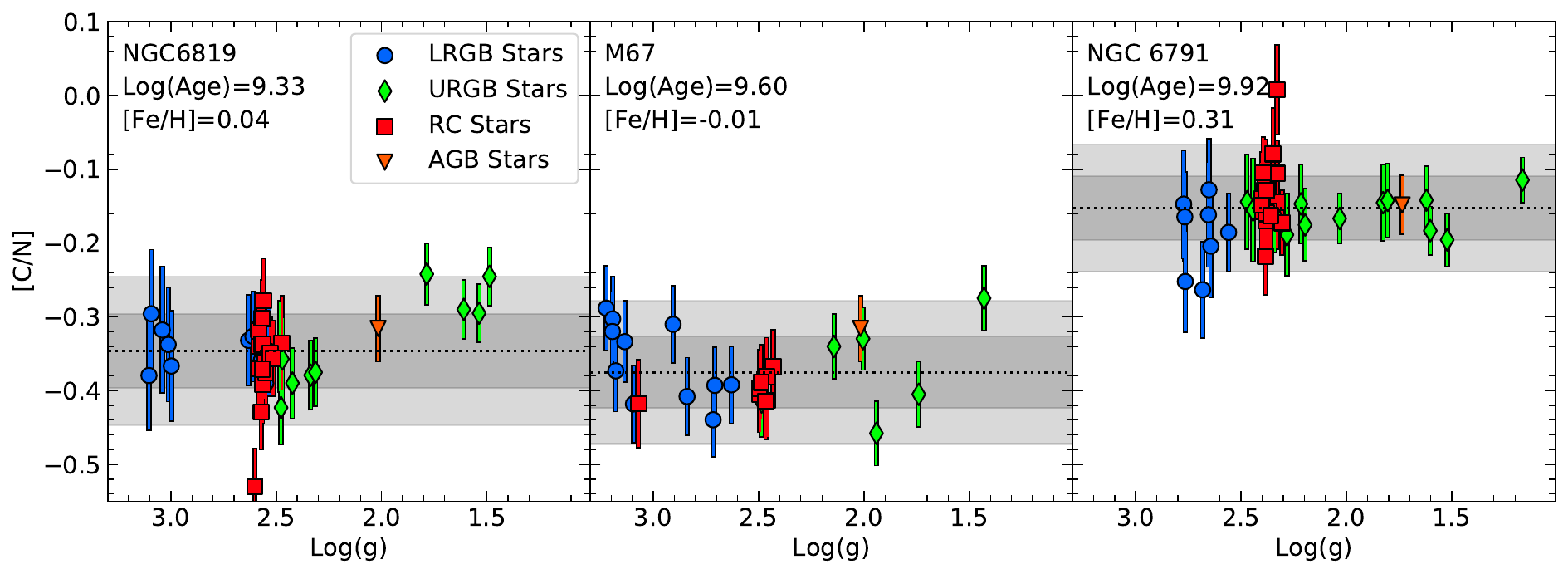}
    \caption{\cn\ versus \logg\ for three different clusters. The dotted line and shaded region show the median \cn\ for the red giants in the clusters, as well as the 1$\sigma$ and 2$\sigma$ deviations from this value. The \cn\ in these clusters does not show strong deviations past the RC stars. NGC 6819 does appear to have an upwards trend in \cn, but this is not beyond the level of scatter expected given the width of the distribution.}
    \label{fig:clusterluminous}
\end{figure*}

Figure \ref{fig:clusterluminous} shows the \cn\ versus \logg\ for all three clusters. NGC 6791 and M67 show consistent \cn\ throughout all 4 stages of evolution. Although there are some offsets in NGC 6819, they are not statistically significant, and they do not appear in the other two clusters. The absence of a clear offset, also supported by \citet{Shetrone_2019} and \citet{Roberts_2024}, is an encouraging sign for \cn\ as a chronometer. This greatly expands the selection of stars we can use in our mapping sample for Section \ref{sec:archeology}.

\subsection{Comparison with Previous Calibrations}\label{val:theory}

\begin{figure}
    \includegraphics[width=0.9\columnwidth]{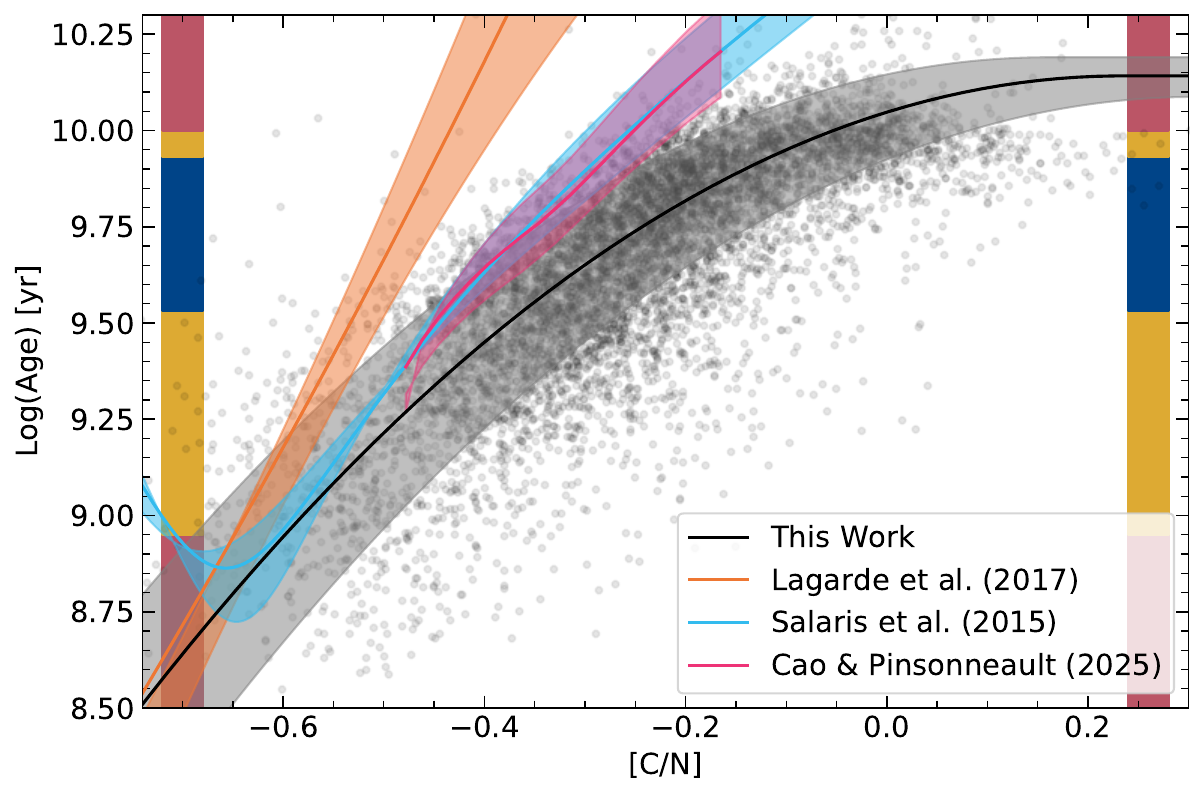}
    \caption{Expected age for a given \cn\ using fits from this work, \citet{Salaris_2015}, \citet{Lagarde_2017}, and \citet{Cao_2025}. The lines shown for each fit are solar \feh\ lines, with the shaded regions showing the changes of 0.4 metallicity in either direction. The lines from \citet{Cao_2025} were limited in age due to concerns about their behavior at high mass. Other plot conventions match those of Figure \ref{fig:empcompare}}
    \label{fig:theocompare}
\end{figure}

\added{Our final comparisons we can make are with the previous calibrations of the \cn-age relationship. First, we compare with three theoretical calibrations:} \cite{Salaris_2015}, who used the BaSTi code \citep{Basti_1, Basti_2} to simulate dredge-up, \cite{Lagarde_2017}, who used the Besançon Galaxy model \citep{Besancon_1, Besancon_2}, a stellar population synthesis model, with input evolutionary tracks produced using STAREVOL \citep{starevol}, and \citet{Cao_2025}, who used the MESA code \citep{MESA_1, MESA_2, MESA_3, MESA_4, MESA_5} to simulate dredge-up and calibrated to the APOKASC-3 stars. Figure \ref{fig:theocompare} mirrors Figure \ref{fig:empcompare}, but shows the theoretical calibrations instead. Here, there is a noticeable difference between the theoretical calibrations and our own. All three theoretical trends predict older ages at a given \cn\ than our fit, but with noticeably lower \cn\ values for a given age. This could be due to over-prediction of mixing or due to a difference in abundance zero-points between their model calibrating stars and the APOGEE catalog, such as discussed in \citet{Cao_2025}.

\begin{figure}
    \includegraphics[width=0.9\columnwidth]{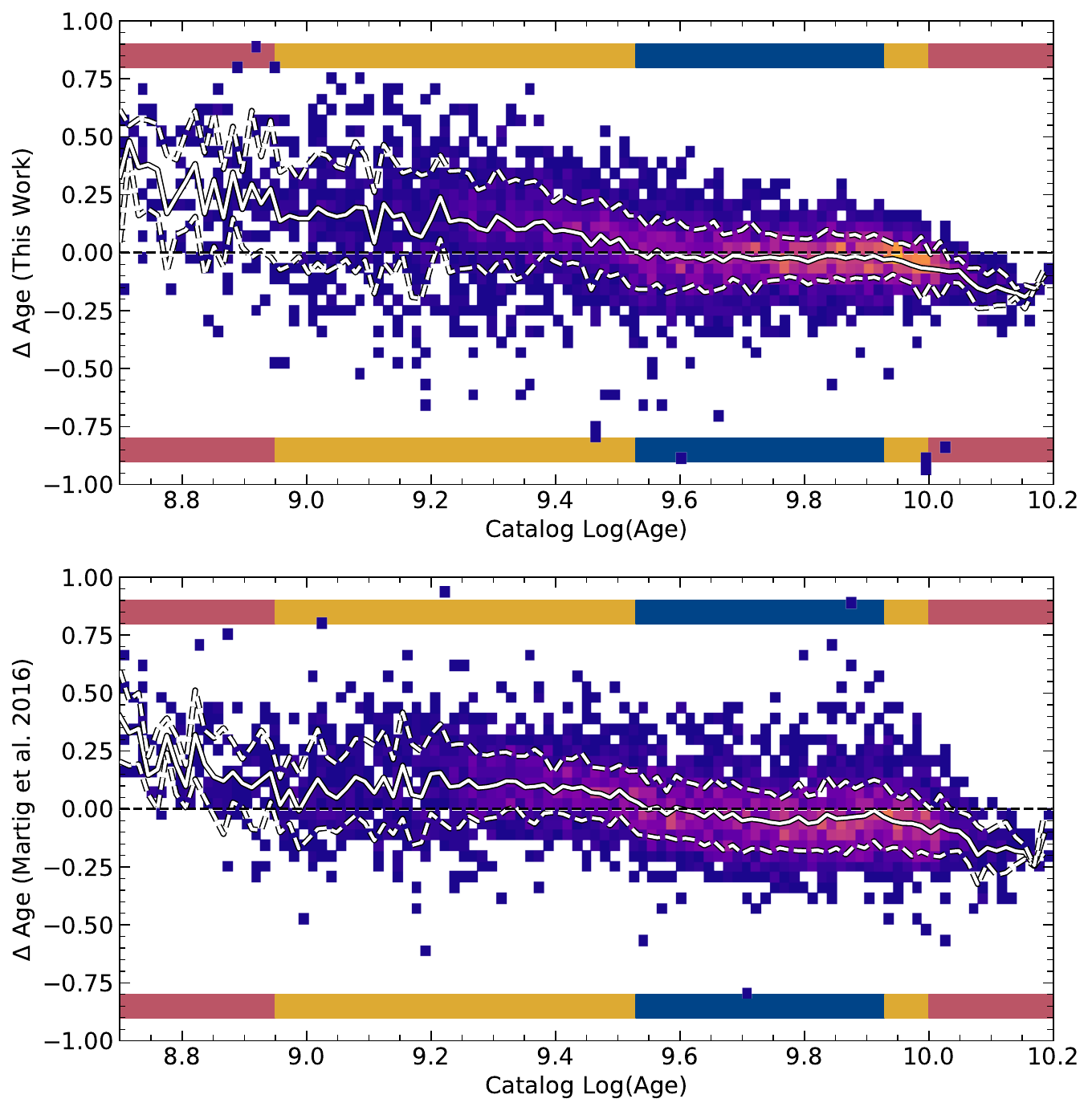}
    \caption{Differences between the \cn\ predicted ages and catalog ages for both this work (top panel) and \citet{Martig_2016} (bottom panel) for the calibration sample stars that have DR12 abundances. The two fits appear similar, but our fit shows lower scatter in the optimal and fair domains.}
    \label{fig:martigcompare}
\end{figure}

\added{Finally, we compare with \citet{Martig_2016}, who calibrated the relationship using APOKASC2 \citep{apokasc2} and APOGEE DR12 \citep{dr12}. As they note, their calibration cannot be directly applied to data outside of DR12 without recalibration, so as a test, we cross-match our calibration sample with APOGEE DR12, and use the DR12 raw abundances for those stars in their fit. This subsample contains 5232 stars and the results of the fits in of their work and this work are shown in Figure \ref{fig:martigcompare}. Structurally, there are only minor differences between our results, which is to be expected given the similarity of our methods. However, the RMS scatter of their results are higher, at 2.14 Gyr, as compared to the 1.64 Gyr of this work. While our fits are structurally quite similar, the enhanced precision of more modern datasets has allowed us to reduce the uncertainty by $\sim25\%$.}

\section{Galactic Archaeology with APOGEE}\label{sec:archeology}

We apply the fits we obtain in Section \ref{sec:fitting} to our mapping sample, as laid out in Section \ref{data:samples}. Stars with ages in the poor region, described in Section \ref{val:recover}, are removed from the sample. After this cut, we are left with 190,643 stars throughout the Galaxy with known ages and surface chemistry.

\begin{figure}
    \centering
    \includegraphics[width=\columnwidth]{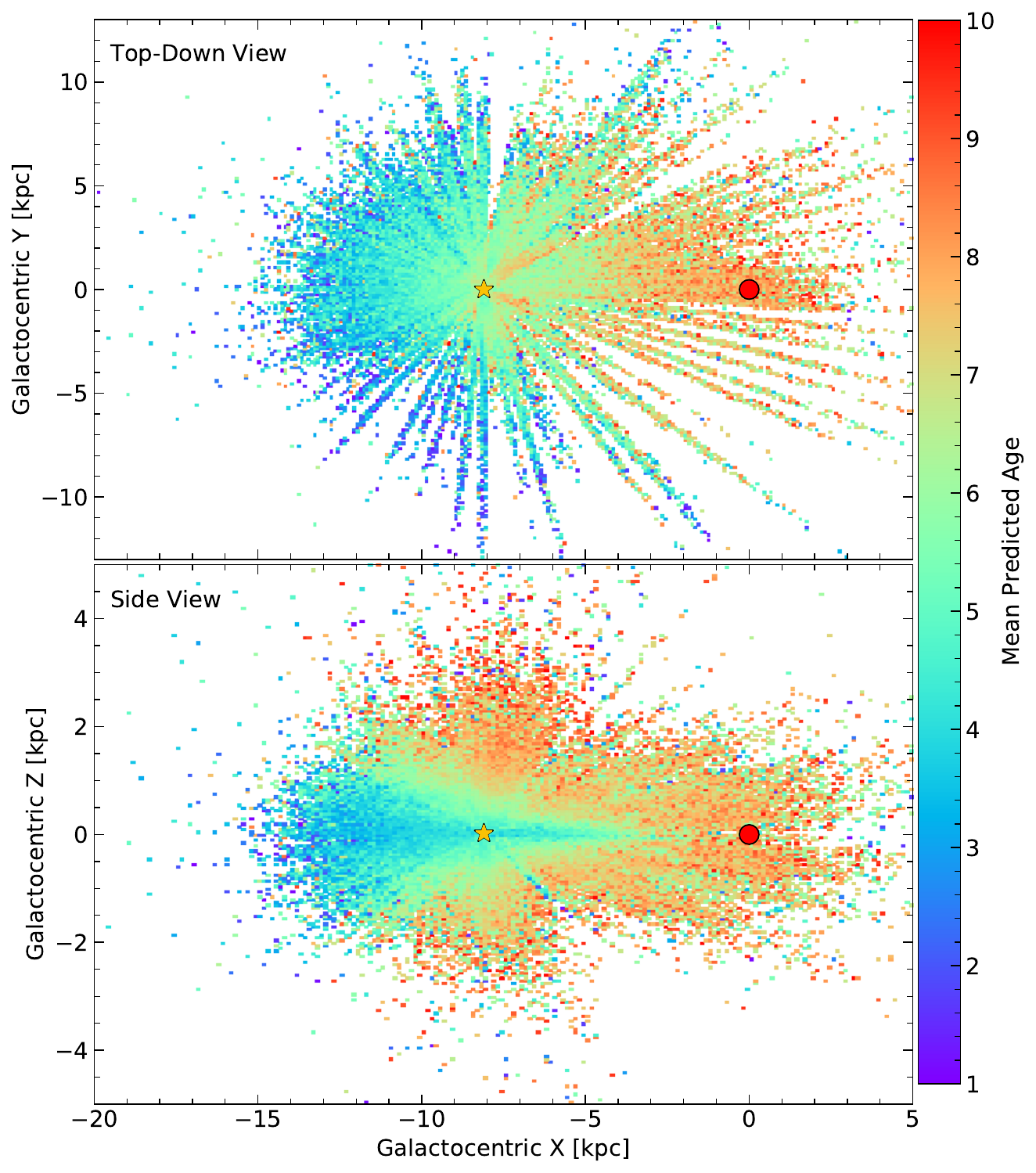}
    \caption{Same plot as Figure \ref{fig:geemapevstate} except now bins are colored by the mean of the predicted ages and all samples are combined}
    \label{fig:agemap}
\end{figure}

Figure \ref{fig:agemap} shows the mapping sample in galactic coordinates like Figure \ref{fig:geemapevstate}, but now colored by median age in each bin rather than bin counts, with all samples combined. We can see how the inner Galaxy is, on average, older than the outer Galaxy, as well as how the youngest stars are primarily centered around the Galactic mid-plane. The young star distribution also flares upwards in the outer Galaxy, as has been observed in other studies \citep{Miglio_2013, Ness_2016, Yu_2021, Leung_2023, Anders_2023}.  We also see an older, inner Galaxy and a younger, outer Galaxy, indicative of inside-out disk formation \citep[e.g.][]{WhiteFrenk_1991, Kauffmann_1996, Bird_2013}. 

\subsection{Locating Outward Migrators}\label{arc:migr}

\added{We wish to reconstruct the chemistry of the Milky Way disk as a function of time. However,} stars are not static within the Milky Way; they can migrate to different Galactic radii \citep[e.g.][]{SellwoodBinney_2002, SchonrichBinney_2009}. In order to use these stars to determine the history of the Galaxy, we need to be able to understand which stars are representative of their current location and which are not. 

Stars can move outwards and inwards, and it is expected that for a single star, the chance to move in one direction is as likely as the other \citep[e.g.][Figure 1]{Johnson_2021}. However, because the inner Galaxy is more densely populated, we are more likely to observe outward migrators than inward migrators in most of the Galaxy, though the exact balance would vary with position. Over much of our domain, outward migrators are expected to be the dominant sample and can be located in a way that inward migrators cannot.

Detailed studies have been conducted of the current gas phase abundance of the Milky Way's interstellar medium and star-forming regions \citep{Braganaca_2019, MendezDelgado_2022}. The Milky Way is richer in metals, specifically oxygen, in the interior than in the exterior at present. Assuming that the ISM has not been significantly more metal-rich in the past than it is now, we can locate these outward migrators via their metallicity. Any star that is significantly more metal-rich than its surrounding environment is likely to have arisen from migration from an interior, more metal-rich environment and migrated to its current position \added{\citep{Lehmann_2024}.}

\added{Given recent studies that propose a potential recent gas infall \citep{Spitoni_2023, Spitoni_2024, Palla_2024}, it is possible that these stars were simply born locally, then had the gas around them diluted. However, for this to be a significant cause of error in this assumption, the infall would need to be quite recent or result in long-lasting depletion, such that the modern gas-phase abundances are significantly lower than older values, which does not seem to be the case \citep{Lian_2022}.}

We flag as an outward migrator any star whose metallicity exceeds the gas-phase metallicity as measured by \citet{MendezDelgado_2022} at its present-day radius. For a conservative selection, we adopt their highest zero-point and require the stars to exceed the gas-phase metallicity by at least one sigma. This process results in the following criterion:

\begin{equation}
    [O/H]_{*}-\sigma[O/H]_{*} > 0.59R_{guide,*}+9.27.
\end{equation}

\begin{figure}
    \centering
    \includegraphics[width=\columnwidth]{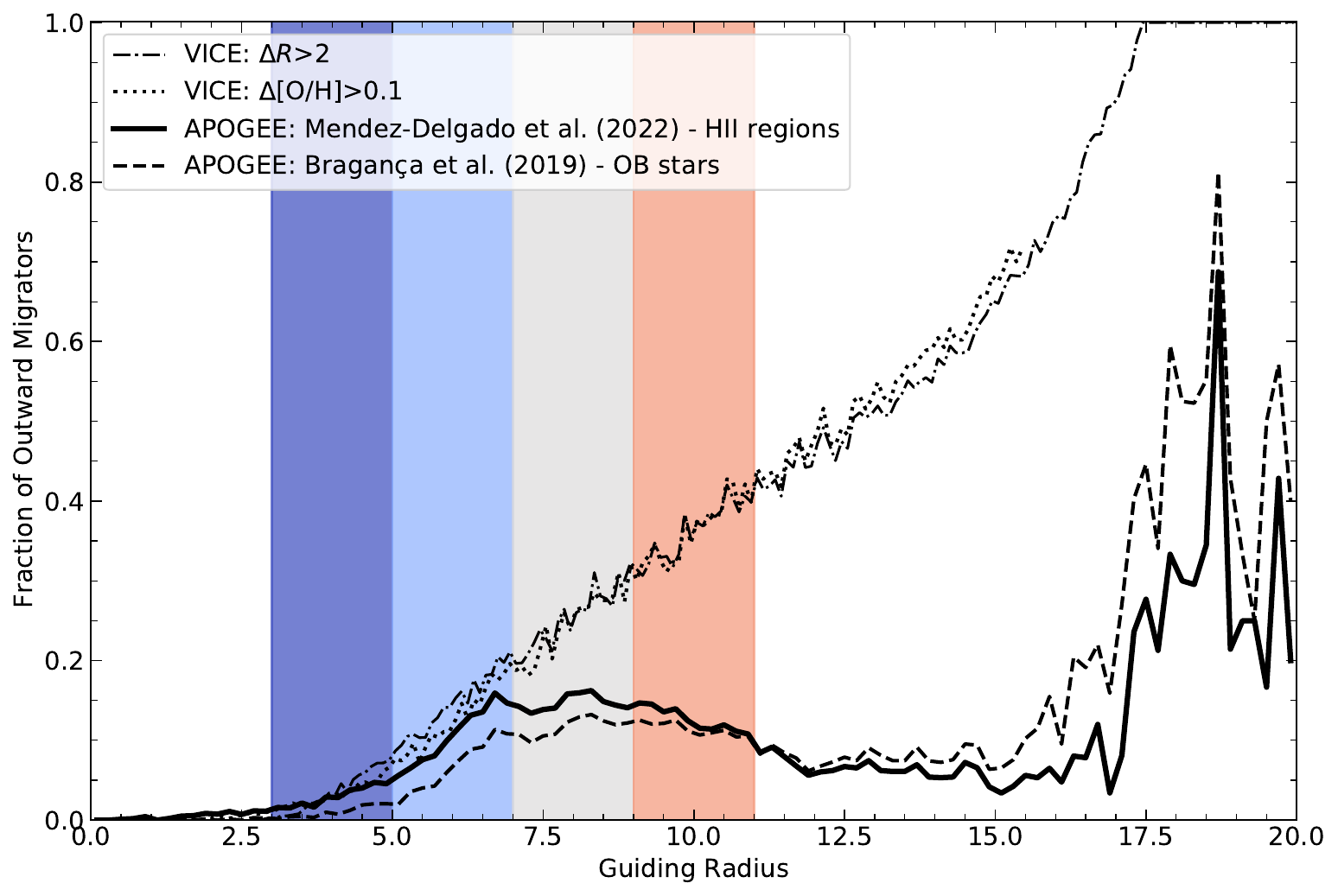}
    \caption{Migrator fractions as defined in Section \ref{arc:migr} versus guiding radius. The two lines labeled VICE are output from our VICE model. The two lines labeled APOGEE come from comparing APOGEE DR17 data with expected radial oxygen gradients, from sources named in the labels. The agreement of the VICE line indicates that using an oxygen cut to find outward migrators is supported by the theory. The agreement of the two APOGEE lines indicates that the pattern we observe is not a consequence of what we use as the baseline.}
    \label{fig:migratorfraction}
\end{figure}

\added{Additionally, because this constraint is a poor match for halo stars, we remove stars with a maximum Z height of 3 or greater were removed.} Applying this criterion results in \added{43,101 out of the total 182,244 stars (23.56\%)} in the mapping sample flagged as migrators. This ratio is a slight overestimate due to our extra-mixing cut removing metal-poor stars in the mapping sample. If we do not remove the stars that have a chance to undergo extra-mixing, then \added{43,676 out of 235,365 (18.56\%)} stars are migrators. The fractions with these stars restored are what we consider in Figure \ref{fig:migratorfraction} and for the rest of this section.

For comparison, we examine the output of a multi-zone galactic chemical evolution model with a prescription for radial migration. We use the Versatile Integrator for Chemical Evolution (VICE; \citep{VICE}), adopting the inside-out star formation history, nucleosynthetic yields, and outflow settings from \citet{Johnson_2021}, and applying the migration prescription described in Appendix C of \citet{Dubay_2024}. Specifically, the distance a model star particle migrates over its lifetime, $\Delta R=R_{\rm final}-R_{\rm form}$, is drawn from a Gaussian of width
\begin{equation}
    \label{eq:migration}
    \sigma_{\rm RM}=2.68\,{\rm kpc} \Big(\frac{\tau}{8\,{\rm Gyr}}\Big)^{0.33} \Big(\frac{R_{\rm form}}{8\,{\rm kpc}}\Big)^{0.61},
\end{equation}
The parameters controlling the migration strength as a function of age and birth radius were tuned to the hydrodynamical simulation {\tt h277} \citep{Christensen_2012}.

Within the model, we flagged migrators in two ways. First, we applied a criterion similar to the one applied to the mapping sample, where stars that were more oxygen-rich than their final environment by over 0.1 dex were flagged as migrators. However, because \added{VICE records the stars' birth radii}, we could also directly count migrators by looking for stars that have moved a certain distance from their birth radius. Because azimuthal abundance variations are of order 0.05-0.1 dex, and the gradient is about 0.06 dex, a minimum of 2 kpc is needed to be sure that a star has a metallicity higher than the ISM in the mapping sample. This also corresponds to the width of the radial bins shown in Figure \ref{fig:migratorfraction}, which are used throughout Section \ref{arc:hist}.

Figure \ref{fig:migratorfraction} shows the fraction of stars that are outward migrators at different radii, both for the APOGEE data and the VICE models. The two VICE model lines track each other quite well, indicating that locating outward migrators with the oxygen condition is effective, at least for a galaxy with a chemical history like that in the model. 

The APOGEE data do not agree with the VICE model. At 6.5 kpc, the two diverge strongly. What is causing the difference between the model trends and the observed trends past 6.5 kpc is unclear. This is not a result of comparing gas phase abundances with stellar abundances, as flagging migrators with a radial gradient derived from OB stars \citep{Braganaca_2019} shows the same discrepancies. 

There are 2 possible causes within the models, which are not mutually exclusive. The first could be that the migration prescription in Equation \ref{eq:migration} is incorrect. In it, the strength of radial migration increases with radius, but an alternate prescription with less growth or weakening of the strength of migration with distance would do much to bridge the gap. Another possible cause could be related to the growth of the disk. The models we ran include star formation at all radii from the start. If we instead limit the outer radii to only begin star formation at later times, then the local stars in the outer galaxy would not have sufficient time to migrate.

However, the trends from VICE are what we would expect to see, where migrators become more common monotonically as we examine regions further out in the Galaxy. In the outer regions, the MW becomes less and less dense, both in stellar populations and in star-forming gas. It would be harder to form stars at 15 kpc than at 5 kpc, and having a large portion of those stars originate from the inner Galaxy is reasonable. The behavior of the APOGEE trends is unexpected, in that around the solar circle, the trend stagnates and decreases by a small amount. \added{It is also unlikely to be caused by a bias in APOGEE selection. APOGEE's target selection for giants only included a blue color cut \citep{Zasowski_2013}, which would remove more metal-poor stars, creating artificially higher migrator fractions. Given that the observed migrator fraction is below expectations, such an explanation does not alleviate the issue.}

\added{Of course, the Milky Way has had a complicated history of merging and cannibalizing other Galaxies. These effects would cause some additional chaos not accounted for in the VICE model. A first-order expectation of a merger would be increased migration during the merger as stars get kicked around. However, once again, the trend we see is counter to that, where the observed counts are far below expected. For a galactic merger to help explain this discrepancy, it would need to either preferentially push stars inward (a direction we cannot capture this way) or deposit a large number of more metal-poor stars into the Galaxy, increasing the number of "non-migrators". However, such an increase in star counts is not expected, given the relative scales of the galaxies the MW has cannibalized.}

Ultimately, we do not know the cause of the discrepancy between these results and believe that it merits future study. For our purposes, we continue to use this locally metal-rich cut to remove outwards migrators for our further analysis. 

\subsection{Milky Way Chemical History}\label{arc:hist}

With our mapping sample now cleaned of the obvious outward migrators, we can use our sample to examine how the chemical trends in the Galaxy have changed over time. We limit the sample to stars with a maximum Z height in their orbit of 0.5 kpc to trace the chemical evolution of the thin disk. The high-alpha thick disk shows a flat age trend \citep{Martig_2016,apokasc3}, but its fractional contribution to the bins varies substantially with radius. Cutting with maximum Z instead of current Z helps avoid this, so our results are not influenced by the fraction of high-alpha stars in the radial bin.

For the remaining stars, we separate them into 1 Gyr-wide age bins and 2 kpc wide guiding radius bins. In each bin, we fit the distribution of \feh\ and \mgh\ with a skew-normal distribution and calculate the mode. The use of mode as our statistic was based on the work of \citet{Johnson_2024} to mitigate the effect of inward migrators and serve as a better proxy for ISM abundance at a given radius and time. The error on the mode was calculated through jackknife resampling with 10 resamples. 

We limit our range to 3-11 kpc. Within 3 kpc, our sample both decreases in size and becomes heavily influenced by the bar, which could introduce notable azimuthal variations that we do not consider. Beyond 11 kpc, the range of expected metallicities of stars extends low enough that the distributions are significantly impacted by our removal of extra-mixing stars. 

\begin{figure*}
    \centering
    \includegraphics[width=2.1\columnwidth]{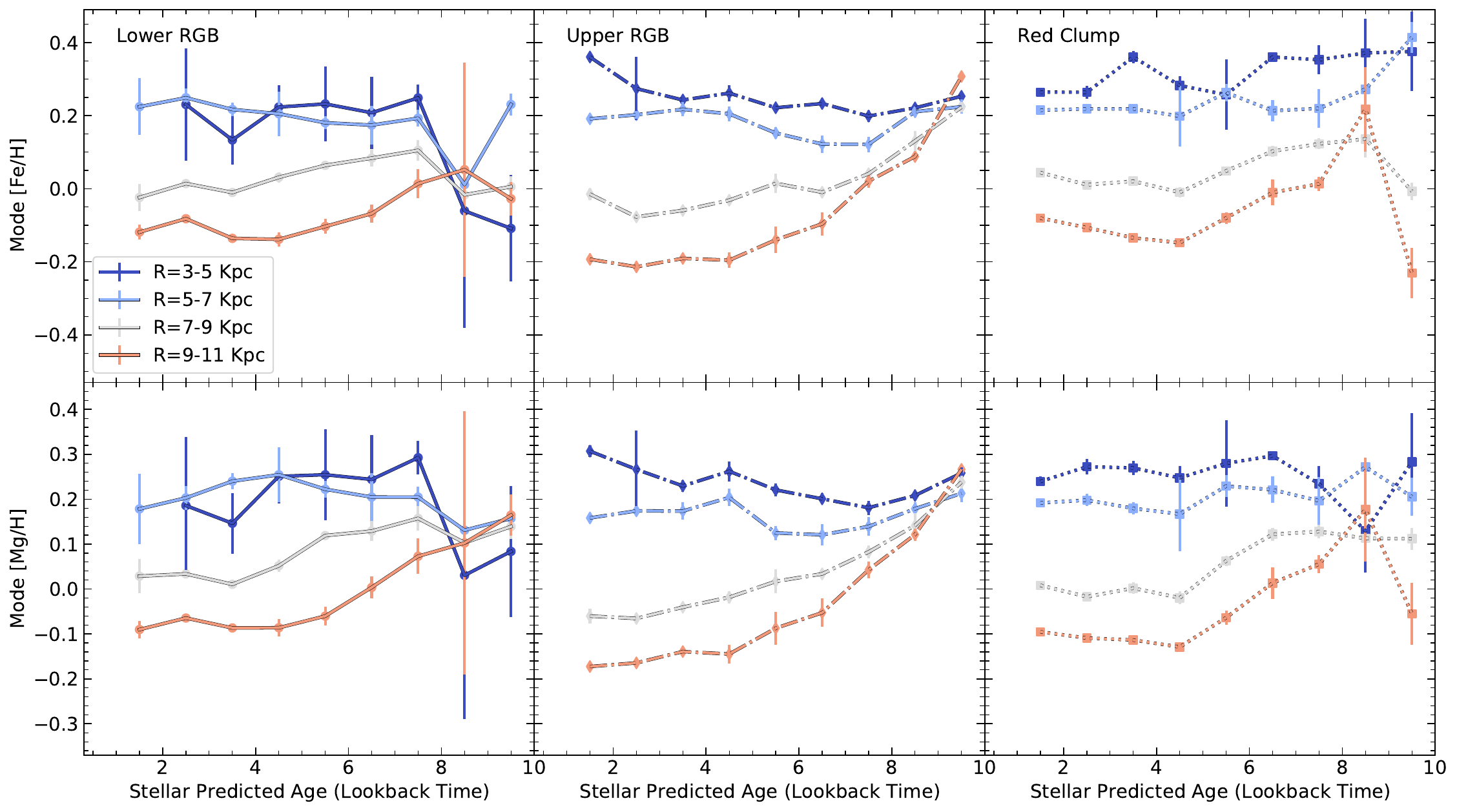}
    \caption{\added{Mode \feh\ (top panels) and \mgh\ (bottom panels) versus stellar age for different guiding radii and evolutionary states. These bins include all stars in the mapping sample, even those flagged as migrators, which is the cause of the seemingly consistent metal-depletion over the last 10 Gyr in the outer radii.}}
    \label{fig:modeabundmigrators}
\end{figure*}

\begin{figure*}
    \centering
    \includegraphics[width=2.1\columnwidth]{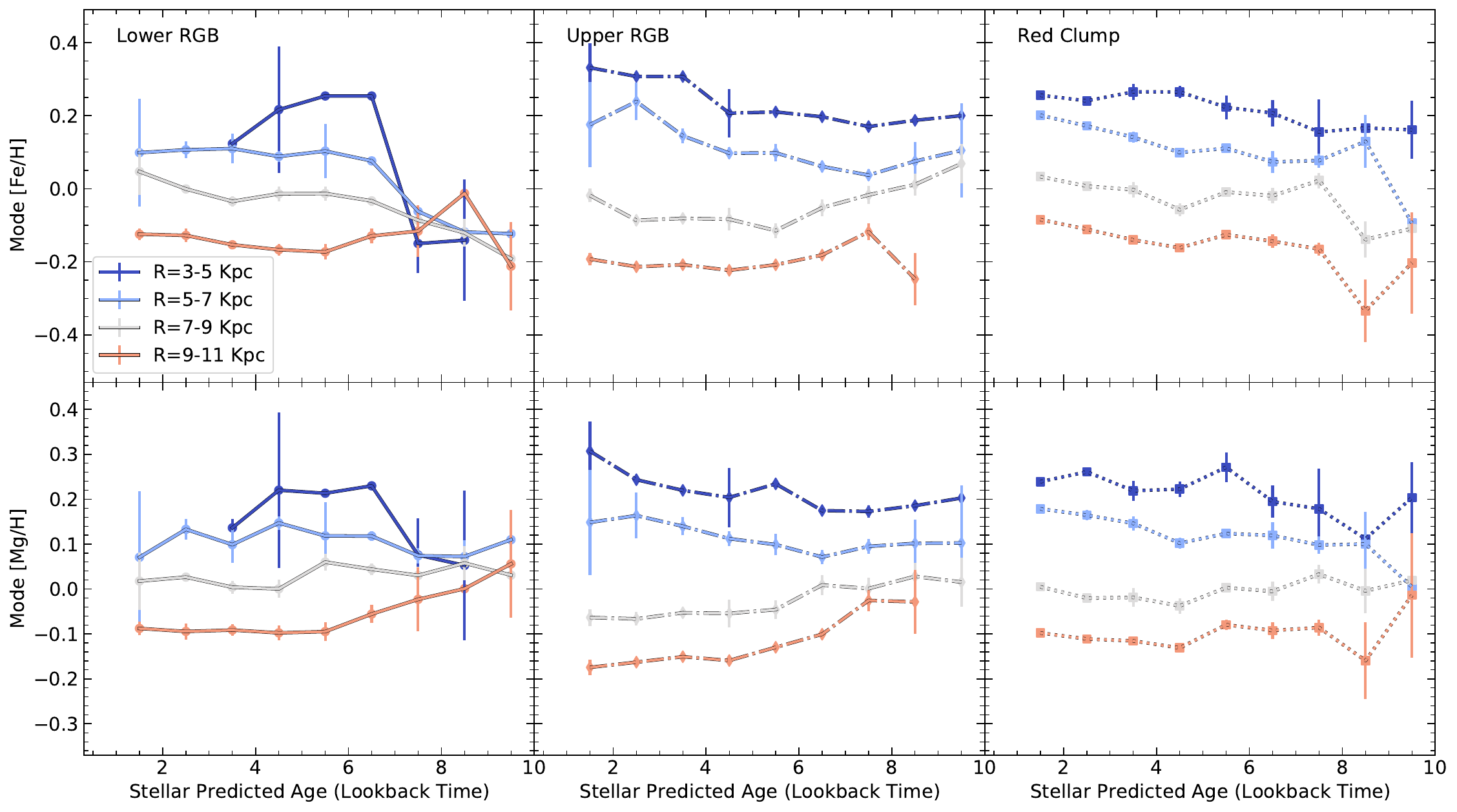}
    \caption{Identical to Figure \ref{fig:modeabundmigrators}, but now removing flagged migrators. All the trends now show weak to no evolution over the last 10 Gyr. The few exceptions, such as rising \mgh\ in the outer Galaxy URGB stars or chaotic inner Galaxy LRGB are more probable sampling issues or remaining migrators rather than actual trends.}
    \label{fig:modeabund}
\end{figure*}

\added{Figures \ref{fig:modeabundmigrators} and \ref{fig:modeabund} show the resulting \feh\ and \mgh\ trends versus stellar age at different radii, including and excluding the stars flagged as migrators, respectively. In Figure \ref{fig:modeabundmigrators}, all samples show a convergence of metallicities in the older bins, at metallicities typical of the solar radius or interior. Older stars have had longer to migrate, and the outer regions of the Galaxy have fewer in-situ stars at those ages, allowing the old migrates stars to strongly affect the sample in those bins. Removing the migrators tells a more believable, though still interesting, story.}

We see very little change in the mode of \feh\ with time. The URGB and RC samples show very weak trends over the last 9-10 Gyr at all radii. The single exception is the 10 kpc radius bin in the URGB, which shows a decrease in \mgh\ over time. This trend is likely due to some remaining outward migrators that could not be fully removed, as our cut was constructed to avoid being overly strict. 

The LRGB paints a slightly different picture, especially at a radius of 4 kpc. However, we believe this difference is largely due to the poor counts of LRGB stars at that radius, \added{(See Figure \ref{fig:geemapevstate}, Left Panel)}. The most striking feature, the jump down between 7 and 8 Gyr, could be explained by inward migrators dominating the sample, as the metallicity is in line with the outer radii for just that bin. Examining the other LRGB lines, the trends are quite similar to the URGB and RC for the last 6 Gyr. Past 6 Gyr, the 6 and 8 kpc lines decrease, which could be an indication of stronger evolution as the Galaxy approaches an equilibrium, but this trend is not observed in the URGB or RC.

We fit each trend with a linear regression and take the median slope across all radii and all evolutionary states. For \feh, the typical enrichment is 0.0176 dex/Gyr. For \mgh, it is practically 0, at 0.0009 dex/Gyr. This is likely due to the aforementioned migrator signal in the outer radii. If we restrict to just examining the inner two radius bins, the typical enrichment is 0.0117 dex/Gyr.

\begin{figure*}
    \centering
    \includegraphics[width=2.1\columnwidth]{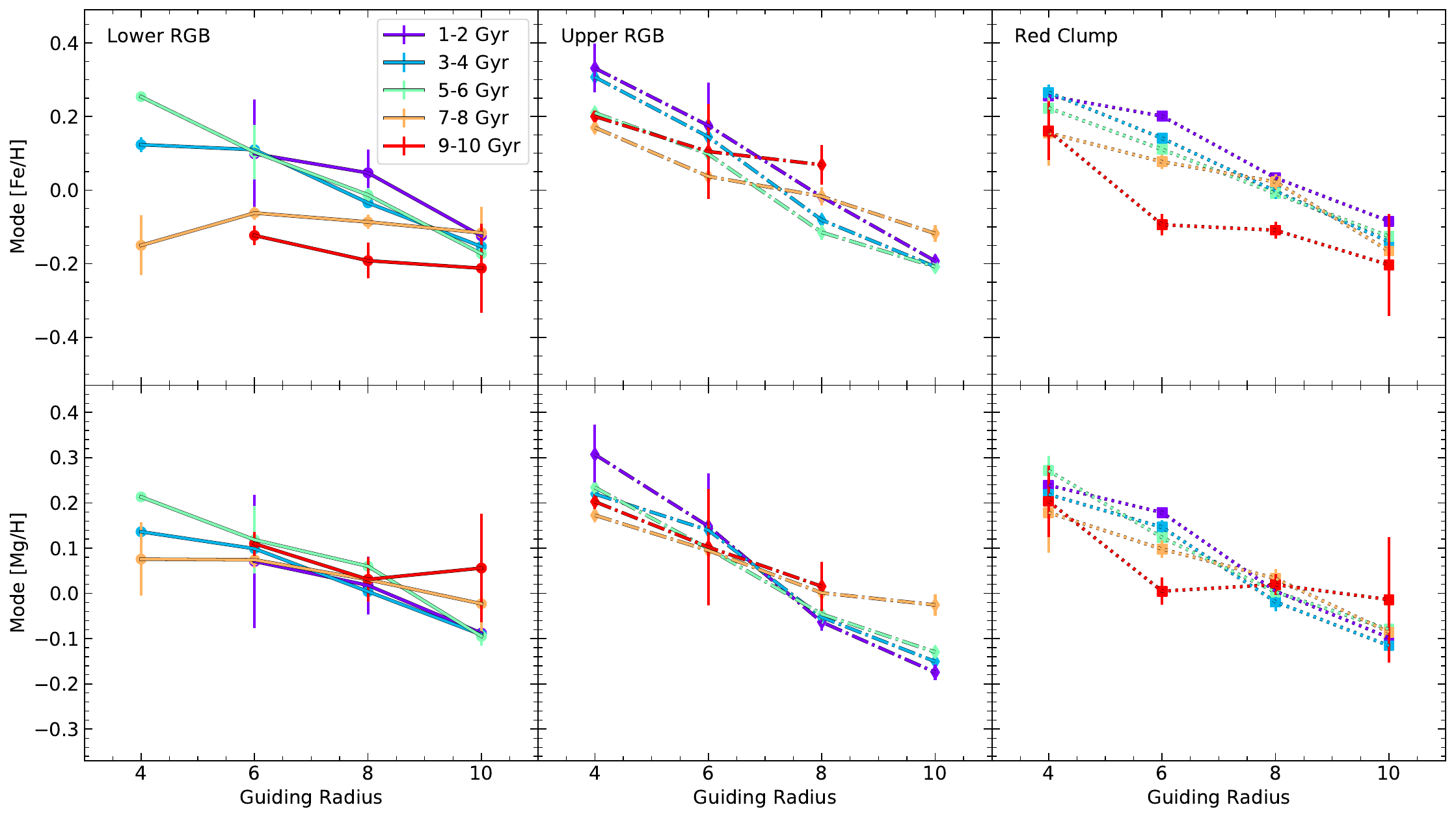}
    \caption{The radial \feh\ and \mgh\ gradients versus at different time bins. Every other time bin has been shown to avoid clutter. The values and error bars used are the same as Figure \ref{fig:modeabund}.}
    \label{fig:gradient}
\end{figure*}

\added{In terms of the evolution of the radial gradient, Figure \ref{fig:gradient} mirrors Figure \ref{fig:modeabund}, but versus radius rather than age. As seen before, the LRGB sample struggled to sample the inner Galaxy at old times, but in the younger domain shows a consistent trend of $\sim$-0.060 dex/kpc. The URGB appears to show a steepening gradient with time (increasing from $\sim$-0.047 dex/kpx in the older bins ago to $\sim$-0.090 dex/kpc in the recent bins), though it is more likely that this is the previously mentioned remaining migrator contamination pushing the older, outer samples to higher metallicities. The RC sample corroborates this, showing a consistent slope of $\sim$-0.063 dex/kpc with a very slowly rising zero-point.}

In order to create these trends, there are two possibilities. First, if migration is significantly faster than previously thought, the result could be stars moving outwards as the ISM is enriched, creating a flat trend when observing the stars. However, given the results in Section \ref{arc:migr}, we find this to be unlikely, as the observed obvious migrators are already less numerous than expected.

A second possible origin for a flat trend in metallicity with age is equilibrium chemical evolution \citep{Chen_2023, Johnson_2024}. GCE models suggest that galaxies evolve toward chemical equilibria where metal production by stars is balanced by hydrogen gained through accretion \citep[e.g.][]{Larson_1974, Weinberg_2017}. However, whether or not the Milky Way reached a state of chemical equilibrium within the lifetime of the Universe is a different question. \citet{Johnson_2024} argued that the Milky Way reached a chemical equilibrium state early in the thin disk epoch based on similar results as our Figure \ref{fig:modeabund}. If their proposed equilibrium scenario is accurate, then the lack of a substantial decline in metallicity across a broad range of ages is a direct consequence thereof. However, the detailed origins of chemical equilibrium remain unknown. \citet{Johnson_2024} demonstrate that outflows ejecting ISM material from the Galactic disk are one possible mechanism, but they did not rule out other possibilities (e.g., radial gas flows; see discussion in their Section 5.1).

\added{Again, this topic is complicated by the merger history of the Milky Way, but the Milky Way has had a relatively quiet merger history for the last few billion years, with the most pronounced merger events taking place at the oldest edge of our range \citep[e.g.][]{Belokurov_2018}. In order to recreate these flat trends through mergers, it would require an ever-escalating series of mergers, as the metal dilution would be accompanied by increased star formation. Though with finer age and spatial resolution than we currently have, some interesting substructures may arise from events like small galactic mergers, which would be on a much smaller scale than the broad equilibrium we observe.}

Our trends showing weak chemical evolution are quite different from previous studies, such as \citet{Lian_2022}. Their results show chemical evolution on the order of 0.2-0.3 dex changes in the last 6 Gyr, and a consistent growth without a radial metallicity gradient longer than 6 Gyr ago. Our differences past 6 Gyr primarily arise due to our mid-plane cut, which removes many of the high-alpha stars. Allowing the high-alpha stars into the sample pulls the inner Galaxy at ages around the age of the thick disk down to comparable metallicities to the outer disk. Our removal of these stars shows the thin disk maintaining a radial gradient for the last 8 Gyr. Additionally, their Figure A1, where they resampled for consistent stellar representation across surface gravities, shows much weaker evolution. Our separation of different evolutionary states accomplishes a similar effect.

\section{Summary and Conclusions}\label{sec:conclusions}

Asteroseismology and spectroscopy are a powerful combination. In this paper, we calibrate a \cn-age relationship for red giants and demonstrate that this relationship can be applied to the vast majority of red giant and red clump stars throughout the Galaxy. Our work takes advantage of precise and accurate asteroseismic data from APOKASC-3, and of our earlier, related work on \cn\ as a diagnostic of stellar mass.

Red giants span a wide range of metallicities, ages, and evolutionary states. It is therefore important to test the circumstances under which \cn\ is a good chronometer. One domain where \cn\ is especially valuable is the luminous giants. In this domain, asteroseismic scaling relations break down, making the masses and ages of these stars difficult to obtain. However, \cn\ can still be measured, and works just as consistently as an age diagnostic for the luminous giants as the less luminous giants. This means that, although it may have less peak precision than asteroseismology, it is still an invaluable tool for exploring the domains where asteroseismology can struggle. Another feature of \cn\ is its insensitivity to mass loss. The \cn-mass correlation is established at the base of the RGB and so is correlated with the birth mass. This gives \cn\ an additional advantage over asteroseismology, which measures the current mass of a star: properly calibrated \cn\ ages do not require specific prescriptions for mass loss. This fact can additionally be exploited to serve as constraints and tests of RGB mass loss prescriptions.

There are domains where \cn\ struggles as a chronometer, however. Primarily, it can only function for stars that have not experienced notable stellar interactions. External alterations to a star's surface chemistry will disrupt the correlation, making \cn\ a misleading chronometer. Given the difficulty of locating stellar merger products, this imposes a difficult limit on the use of \cn. Even for single stars with \feh\ below $-0.4$, extra mixing complicates the interpretation of \cn\ as an age diagnostic in the RC and on the URGB. Our results cannot be applied to this metal-poor regime due to this. Additionally, the range of ages that can be recovered by \cn\ has limits on both the young and old sides. Ages younger than 1 Gyr and older than 10 Gyr are insensitive to \cn\ and cannot be recovered. Recovering ages from $1-3.5$ Gyr as well as $8.5-10$ Gyr is challenging. There are issues with accuracy, but not beyond the level of precision. Even within the most optimal domain, \cn\ ages still boast notable uncertainties arising primarily from differences in birth \cn\ values. Our results show consistent 1.64 Gyr scatter across the entire $1-10$ Gyr domain. 

Regardless, \cn\ still provides a valuable glimpse of a star's age across a wide range of conditions. To exploit this, we applied our method to the full APOGEE catalog, carefully selecting stars in the domain where \cn\ is a good clock. We applied kinematic and chemical cuts to isolate populations of stars that we believe to be migrators. We noticed less migration than expected from models in the outer galaxy, which we believe is worth follow-up examination. We then used our age relationship to predict ages for the remaining stars, which should reflect the birth abundance trends across space and time in the Galaxy. We showed empirical evidence of very weak chemical evolution in the Milky Way disk ($\sim$0.01-0.015 dex/Gyr), much weaker than typically assumed. 

The future of this topic will be quite exciting. Applications to larger spectroscopic samples, such as Milky Way Mapper, or combinations with other datasets that include C and N, could expand the reach of these age calibrations. Combining with other larger sets of asteroseismic data, such as TESS \citep{TESS} or the upcoming Roman Galactic Bulge Time Domain Survey \citep[GTBDS,][]{ROMAN}, could also prove quite valuable for enhancing the precision of \cn\ based ages. The use of \cn\ in conjunction with other methods will also allow for all the methods to help alleviate the drawbacks of the others. Further work to quantify the strength of extra mixing effects on the RGB could also allow \cn\ calibrations to apply to the metal-poor domain that we could not capture in this work.

\section*{Acknowledgments}
We would like to acknowledge the OSU "Stars Group" as well as the SDSS-V "Stellar Ages Working Group" and "Galactic Genesis Working Group" for enlightening discussions and comments. \added{We would also like to thank both our reviewers for their insightful comments and help in making this paper clear and accessible.}

J. R., M.H. P., and J.A.J. acknowledge support from NASA ADAP grant 80NSSC24K0637. J.A.J. and L.O.D. acknowledge support from NSF grant AST-2307621. J.W.J. acknowledges funding from a Carnegie Theoretical Astrophysics Center postdoctoral fellowship.

Funding for the Sloan Digital Sky Survey IV has been provided by the Alfred P. Sloan Foundation, the U.S. Department of Energy Office of Science, and the Participating Institutions. SDSS acknowledges support and resources from the Center for High-Performance Computing at the University of Utah. The SDSS website is www.sdss4.org.

SDSS is managed by the Astrophysical Research Consortium for the Participating Institutions of the SDSS Collaboration including the Brazilian Participation Group, the Carnegie Institution for Science, Carnegie Mellon University, Center for Astrophysics | Harvard \& Smithsonian (CfA), the Chilean Participation Group, the French Participation Group, Instituto de Astrofísica de Canarias, The Johns Hopkins University, Kavli Institute for the Physics and Mathematics of the Universe (IPMU) / University of Tokyo, the Korean Participation Group, Lawrence Berkeley National Laboratory, Leibniz Institut für Astrophysik Potsdam (AIP), Max-Planck-Institut für Astronomie (MPIA Heidelberg), Max-Planck-Institut für Astrophysik (MPA Garching), Max-Planck-Institut für Extraterrestrische Physik (MPE), National Astronomical Observatories of China, New Mexico State University, New York University, University of Notre Dame, Observatório Nacional / MCTI, The Ohio State University, Pennsylvania State University, Shanghai Astronomical Observatory, United Kingdom Participation Group, Universidad Nacional Autónoma de México, University of Arizona, University of Colorado Boulder, University of Oxford, University of Portsmouth, University of Utah, University of Virginia, University of Washington, University of Wisconsin, Vanderbilt University, and Yale University.

\software{astropy \citep{astropy2013,astropy2018}, SciPy \citep{scipy}, NumPy \citep{numpy}, Matplotlib \citep{matplotlib}, VICE \citep{VICE}}

\bibliography{biblio}{}
\bibliographystyle{aasjournal}

\end{document}